\documentclass[journal,twoside,twocolumn,letterpaper]{IEEEtran} 
\usepackage{amsmath,amssymb,mathrsfs,amscd,latexsym,dsfont}
\usepackage{color,comment}
\usepackage{enumerate}
\usepackage{cite}
\usepackage{graphicx}
\usepackage[dvipsnames]{xcolor}
\usepackage{amsfonts}
\usepackage{psfrag}
\usepackage[utf8]{inputenc}
\usepackage[T1]{fontenc}

\usepackage{subcaption}

\newcommand{\q}{\boldsymbol{q}}
\newcommand{\n}{\boldsymbol{n}}
\newcommand{\s}{\boldsymbol{s}}

\newcommand{\MI}{I_{X,Y}}
\newcommand{\E}{\boldsymbol{E}}
\newcommand{\N}{\boldsymbol{N}}
\newcommand{\xbt}{\boldsymbol{x}(t)} 	
\newcommand{\ybt}{\boldsymbol{y}(t)}	
\newcommand{\alphaproof}{\alpha}        

\newcommand{\Var}{\mathrm{Var}}

\DeclareMathOperator{\sech}{sech}

\DeclareMathOperator{\var}{Var}
\DeclareMathOperator{\natlog}{\log}

\newcommand{\coa}{\hat{a}}
\newcommand{\cox}{\hat{x}}
\newcommand{\coaN}{\hat{a}_{\textnormal{noise}}}
\newcommand{\coaS}{\hat{a}_{\textnormal{inter}}}

\IEEEoverridecommandlockouts

\newtheorem{theorem}{Theorem}
\newtheorem{example}{Example}

\newtheorem{lemma}[theorem]{Lemma}

\title{Capacity Lower Bounds of the Noncentral Chi-Channel with Applications to Soliton Amplitude Modulation}
\author{Nikita~A.~Shevchenko, Stanislav~A.~Derevyanko, Jaroslaw~E.~Prilepsky, Alex~Alvarado,  Polina~Bayvel, and Sergei~K.~Turitsyn
\thanks{N.~A.~Shevchenko and P.~Bayvel are with the Optical Networks Group, Department of Electronic \& Electrical Engineering, University College London, London WC1E 7JE, UK (email: mykyta.shevchenko.13@ucl.ac.uk).}
\thanks{S.~A.~Derevyanko is with the Department of Electrical and Computer Engineering, Ben-Gurion University of the Negev, Beer Sheva 84105, Israel (email: stasd@bgu.ac.il).}
\thanks{J.~E.~Prilepsky and S.~K.~Turitsyn are with the Aston Institute of Photonic Technologies, Aston University, Birmingham B4 7ET, UK (email: y.prylepskiy1@aston.ac.uk).}
\thanks{A.~Alvarado is with the Signal Processing Systems Group, Department of Electrical Engineering, Eindhoven University of
Technology (TU/e), Eindhoven, 5600 MB, The Netherlands (email: alex.alvarado@ieee.org)
}
	
\thanks{Research supported by the Engineering and Physical Sciences Research Council (EPSRC) project UNLOC (EP/J017582/1), by the Netherlands Organisation for Scientific Research (NWO) via the VIDI Grant ICONIC (project number 15685), and a UCL Graduate Research Scholarship (GRS).}
}

\begin{document}

\maketitle

\begin{abstract}
The channel law for amplitude-modulated solitons transmitted through a nonlinear optical fibre with ideal distributed amplification and a receiver based on the nonlinear Fourier transform is a noncentral chi-distribution with $2n$ degrees of freedom, where $n=2$ and $n=3$ correspond to the single- and dual-polarisation cases, respectively. In this paper, we study capacity lower bounds of this channel under an average power constraint in bits per channel use. We develop an asymptotic semi-analytic approximation for a capacity lower bound for arbitrary $n$ and a Rayleigh input distribution. It is shown that this lower bound grows logarithmically with signal-to-noise ratio (SNR), independently of the value of $n$. Numerical results for other continuous input distributions are also provided. A half-Gaussian input distribution is shown to give larger rates than a Rayleigh input distribution for $n=1,2,3$. At an SNR of $25$~dB, the best lower bounds we developed are approximately $3.68$~bit per channel use. The practically relevant case of amplitude shift-keying (ASK) constellations is also numerically analysed. For the same SNR of $25$~dB, a $16$-ASK constellation yields a rate of approximately $3.45$~bit per channel use.
\end{abstract}


\section{Introduction}

Optical fibre transmission systems carrying the overwhelming bulk of the world's telecommunication traffic have undergone a long process of increasing engineering complexity and sophistication \cite{cd13,w15,bmx16}. However, the key \emph{physical effects} affecting the performance of these systems  remain largely the same. These are: attenuation, chromatic dispersion, fibre nonlinearity due to the optical Kerr effect, and optical noise. Although the bandwidth of optical fibre transmission systems is large, these systems are ultimately band-limited. This bandwidth limitation combined with the ever-growing demand for data rates is expected to result in a so-called ``capacity crunch'' \cite{r10}, which caps the rate increase of error-free data transmission \cite{r10,ms01,ekw10,etr13}. Designing spectrally-efficient transmission systems is therefore a key challenge for future optical fibre transmission systems.

The channel model used in optical communication that includes all three above-mentioned key effects for two states of polarisation is the so-called Manakov equation (ME) \cite[eq. (1.26)]{etr13}, \cite[Sec. 10.3.1]{hk95}. The ME describes the propagation of the optical field for systems employing polarisation division multiplexing. The ME therefore generalises the popular scalar nonlinear Schr\"odinger equation (NSE) \cite{ekw10,Iannone,etr13,hk95}, used for single-polarisation systems. In both models, the evolution of the optical field along the fibre is represented by a nonlinear partial differential equation with complex additive Gaussian noise.\footnote{The precise mathematical expressions for both channel models are given in Sec.~\ref{Sec:CT.Master}.} The accumulated nonlinear interaction between the signal and the noise makes the analysis of the resulting channel model a very difficult problem. As recently discussed in, e.g., \cite[Sec.~1]{agrell16}, \cite{fs15}, \cite{sf17}, exact channel capacity results for fibre optical systems are scarce, and many aspects related to this problem remain open.

Until recently, the common belief among some researchers in the field of optical communication was that nonlinearity was always a nuisance that necessarily degrades the system performance. This led to the assumption that the capacity of the optical channel had a peaky behaviour when plotted as a function of the transmit power\footnote{However, nondecaying bounds can be found in the literature, e.g., in \cite{agrell16,tdy03} (lower bounds) and \cite{kyk15,ykk15} (upper bounds).}. Partially motivated by the idea of improving the data rates in optical fibre links, a multitude of nonlinearity compensation methods have been proposed (see, e.g., \cite{d14,r16,Millar,Ip,Du,Liu}), each resulting in different discrete-time channel models. Recently, a paradigm-shifting approach for overcoming the effects of nonlinearity has been receiving increased attention. This approach relies on the fact that both the ME and NSE in the absence of losses and noise are \emph{exactly integrable} \cite{Zakharov,m74}.

One of the consequences of integrability is that the signal evolution can be represented using nonlinear normal modes. While the pulse propagation in the ME and NSE is nonlinear, the evolution of these nonlinear modes in the so-called nonlinear spectral domain is essentially linear \cite{yk14-1}, \cite{o17}. The decomposition of the waveform into the nonlinear modes (and the reciprocal operation) is often referred to as nonlinear Fourier transform (NFT), due to its similarity with the application of the conventional Fourier decomposition in linear systems \cite{akns74}.\footnote{In mathematics and physics literature, the name \emph{inverse scattering transform} method for the NFT is more commonly used.} The linear propagation of the nonlinear modes implies that the nonlinear cross-talk in the NFT domain is theoretically absent, an idea exploited in the so-called \emph{nonlinear frequency division multiplexing} \cite{yk14-1,yy17}. In this method, the nonlinear interference can be greatly suppressed by assigning users different ranges in the nonlinear spectrum, instead of multiplexing them using the conventional Fourier domain.

Integrability (and the general ideas based around NFT) has also lead to several nonlinearity compensation, transmission and coding schemes \cite{yk14-2,yk14-3,tt13,pdb14,mm15,b15,b15-1,dhg15,mtm15,m15,lpp16}. These can be seen as a generalisation of soliton-based communications \cite{hk95}, \cite[Chapter 5]{Iannone,mg}, which 
follow the pioneering work by Hasegawa and Nyu \cite{hn93}, and where only the discrete eigenvalues were used for communication. The development of efficient and numerically stable algorithms has also attracted a lot of attention \cite{wp15}. Furthermore, there have also been a number of experimental demonstrations and assessments for different NFT-based systems \cite{b15,b15-1,dhg15,mtm15,m15,lpp16}. However, for systems governed by the ME, the only results available come from the recent theoretical work of Maruta and Matsuda \cite{mm15}.

Two nonlinear spectra (types of nonlinear modes) exist in the NSE and the ME. The first one is the so-called continuous spectrum, which is the exact nonlinear analogue of the familiar linear FT, inasmuch as its evolution in an optical fibre is exactly equivalent to that of the linear spectrum under the action of the chromatic dispersion and the energy contained in the continuous spectrum is related to that in the time domain by a modified Parseval equality \cite{pdb14,akns74}. The unique feature of the NFT is, however, that apart from the continuous spectrum, it can support a set of \emph{discrete} eigenvalues (the nondispersive part of the solution). In the time domain, these eigenvalues correspond to stable localised multi-soliton waveforms immune to both dispersion and nonlinearity \cite{hk95}. The spectral efficiency of the multiple-eigenvalue encoding schemes is an area actively explored at the moment \cite{yk14-3,hky16,km16}. Multi-soliton transmission has also received increased attention in recent years, see, e.g., \cite{yrft14} and \cite{gui2017} and references therein. Finding the capacity of the multi-eigenvalue-based systems in the presence of in-line noise that breaks integrability still remains an open research problem. If only a single eigenvalue per time slot is used, the problem is equivalent to a well-known time-domain amplitude-modulated soliton transmission system\footnote{Since the imaginary part of a single discrete eigenvalue is proportional to the soliton amplitude.}. In this paper, we consider this simple set-up, where a single eigenvalue is transmitted in every time slot. The obtained results are applicable not only to classical soliton communication systems, but also to the novel area of the eigenvalue communications.

Although the set-up we consider in this paper is one of the simplest ones, its channel capacity is still unknown. Furthermore, the only results available in the literature \cite{yk14-3,mfs12,hky16,km16,zhch15,zhch16,spd15} are exclusively for the NSE, leaving the ME completely unexplored. In particular, previous results include those by Meron et al. \cite{mfs12}, who recognised that mutual information (MI) in a nonlinear integrable channel can (and should) be evaluated through the statistics of the nonlinear spectrum, i.e., via the channel defined in the NFT domain. Using a Gaussian scalar model for the amplitude evolution with in-line noise, a lower bound on the MI and capacity of a single-soliton transmission system was presented. The case of two and more solitons per one time slot was also analysed, where data rate gains of the continuous soliton modulation versus an on-off-keying (OOK) system were also shown. A bit-error rate analysis for the case of two interacting solitons has been presented in \cite{fst2001}. The derivations presented there, however, cannot be used straightforwardly for information theoretic analysis. Yousefi and Kschischang \cite{yk14-3} addressed the question of achievable spectral efficiency for single- and multi-eigenvalue transmission systems using a Gaussian model for the nonlinear spectrum evolution. Some results on the continuous spectrum modulation were also presented. Later in \cite{hky16}, the spectral efficiency of a multi-eigenvalue transmission system was studied in more detail. In \cite{km16}, the same problem was studied by considering the correlation functions of the spectral data obtained in the quasi-classical limit of large number of eigenvalues. Achievable information rates for multi-eigenvalue transmission systems utilising all four degrees of freedom of each scalar soliton in NSE were analytically obtained in \cite{zhch16}. These results were obtained within the framework of a Gaussian noise model provided in \cite{yk14-3, zhch15} (non-Gaussian models have been presented in \cite{dty03,dty05}) and assuming a continuous uniform input distribution subject to peak power constraints. The spectral efficiency for the NFT continuous spectrum modulation was considered in \cite{dpt16,yy16,ts17}. Periodic NFT methods have been recently investigated in \cite{kpl17}.

In \cite{spd15}, we used a non-Gaussian model for the evolution of a single soliton amplitude and the NSE. Our results showed that a lower bound for the capacity per channel use of such a model grows unbounded with the effective signal-to-noise ratio (SNR). In this paper, we generalise and extend our results in \cite{spd15} to the ME. To this end, we use perturbation-based channel laws for soliton amplitudes previously reported in \cite{dty03,dty05} (for the NSE) and \cite{dpy06} (for the ME). Both channel laws are a noncentral chi ($\chi$) distribution with $2n$ degrees of freedom, where $n=2$ and $n=3$ correspond to the NSE and ME, respectively. Motivated by the similarity of the channel models mentioned above, in this paper we study asymptotic lower bound approximations on the capacity (in bit per channel use) of a general noncentral chi-channel arbitrary (even) number of degrees of freedom. To the best of our knowledge, this has not been previously reported in the literature. Similar models, however, do appear in the study of noise-driven coupled nonlinear oscillators \cite{pd05}.

The first contribution of this paper is to numerically obtain lower bounds for the channel capacity for three continuous input distributions, as well as for amplitude shift-keying (ASK) constellations with discrete number of constellation points. For all the continuous inputs, the lower bounds are shown to be nondecreasing functions of the SNR under an average power constraint. The second contribution of this paper is to provide an asymptotic closed-form expression for the MI of the noncentral chi-channel with a arbitrary (even) number of degrees of freedom. This asymptotic expression shows that the MI grows unbounded and at the same rate, independently of the number of degrees of freedom.

\section{Continuous-time Channel Model}\label{Sec:CT}

\subsection{The Propagation Equations}\label{Sec:CT.Master}

The propagation of light in optical fibres in the presence of amplified spontaneous emission (ASE) noise can be described by a stochastic partial differential equation which captures the effects of chromatic dispersion, nonlinear polarisation mode dispersion, optical Kerr effect, and the generation of ASE noise from the optical amplification process. Throughout this paper we assume that the fibre loss is continuously compensated along the fibre by means of (ideal) distributed Raman amplification (DRA)\cite{QL1,QL2}. In this work we consider the propagation of a slowly varying 2-component envelope $\E(\ell,\tau) = [E_1(\ell,\tau), \, E_2(\ell,\tau)]\in\mathbb{C}^2$ over a nonlinear birefringent optical fibre, where $\tau$ and $\ell$ represent time and propagation distance, respectively. Our model also includes the 2-component ASE noise $\boldsymbol{N}(\ell,\tau)=[N_1(\ell,\tau), \,N_2(\ell,\tau)]$ due to the DRA. We also assume a uniform change of polarised state on the Poincaré sphere \cite{wmc91}. 

The resulting lossless ME is then given by \cite[eq.~(1.26)]{etr13},\cite[Sec. 10.3.1]{hk95},\cite{lk98,dpy06}\footnote{Throughout this paper, vectors are denoted by boldface symbols $\boldsymbol{x}=[x_1,x_2,x_3,...]$, while scalars are denoted by nonboldface symbols. The scalar product is denoted by $\langle \cdot \,, \cdot \rangle$, and over-bar denotes complex conjugation. The Euclidean norm is denoted by $\left\|\boldsymbol{x}\right\|^2 \triangleq \left|x_1\right|^2+\left|x_2\right|^2+...$. The partial derivatives in the partial differential equations are expressed as subscripts, e.g., $\E_{\ell}\triangleq\frac{\partial \E}{\partial \ell}$, $\E_{\tau\tau}\triangleq\frac{\partial^2 \E}{\partial \tau^2}$, etc. The imaginary unit is denoted by $\imath\triangleq\sqrt{-1}$.}
\begin{equation}\label{Manakov.r.w.u}
\imath \E_{\ell} - \frac{\beta_{2}}{2} \, \E_{\tau\tau} + \frac{8 \gamma}{9} \langle \E \,, \bar{\E}\rangle \, \E = \N(\ell,\tau),
\end{equation}
where the retarded time $\tau$ is measured in the reference frame moving with the optical pulse average group velocity, $\E \equiv \E(\ell,\tau)$ represents the slowly varying 2-component envelope of electric field, $\beta_{2}$ is the group velocity dispersion coefficient characterising the  chromatic dispersion, and $\gamma$ is the fibre nonlinearity coefficient. The pre-factor $8/9$ in \eqref{Manakov.r.w.u} comes from the averaging of the fast polarisation rotation \cite[Sec. 10.3.1]{hk95}, \cite{wmc91}. For simplicity we will further work with the effective averaged nonlinear coefficient $\gamma^{\ast} \triangleq  8\gamma/9$ when addressing the ME. In the case of a single polarisation state, the propagation equation above reduces to the lossless generalised \emph{scalar} NSE \cite{ekw10,Iannone}
\begin{equation}\label{NLSE.r.w.u}
\imath E_{\ell} - \frac{\beta_{2}}{2} \, E_{\tau\tau} + \gamma \left|E\right|^{2} E = N(\ell,\tau).
\end{equation}

In this paper we consider the case of anomalous dispersion ($\beta_{2} < 0$), i.e., the \emph{focusing} case. In this case,  both the ME in \eqref{Manakov.r.w.u} and the NSE in \eqref{NLSE.r.w.u} permit \emph{bright} soliton solutions (``particle-like waves''), which will be discussed in more detail in Sec.~\ref{Sec:CT.Solitons}.

It is customary to re-scale \eqref{Manakov.r.w.u} to dimensionless units. We shall use the following normalisation: The power will be measured in units of $P_0=1$~mW since it is a typical power level used in optical communications. The normalised (dimensionless) field then becomes $\q=\E/\sqrt{P_0}$. For the distance and time, we define the dimensionless variables $z$ and $t$ as $z=\ell/\ell_0$ and $t=\tau/\tau_0$, where
\begin{equation}\label{norm_scale}
\ell_0=(\gamma^* P_0)^{-1}, \quad \tau_0=
\sqrt{\ell_0 |\beta_2|}=\sqrt{\frac{|\beta_2|}{\gamma^{*} P_0}}.
\end{equation}
For the scalar case \eqref{NLSE.r.w.u}, we use the same normalisation but we replace $\gamma^*$ by $\gamma$. Then, the resulting ME reads \begin{equation}\label{Manakov.au}
\imath \q_{z}+\frac{1}{2} \,\q_{t t}+\langle \q \,, \bar{\q}\rangle \,\q = \n(z,t),
\end{equation}
while the NSE becomes
\begin{equation}\label{NLSE.au}
\imath q_{z}+\frac{1}{2} \,q_{t t}+ \left|q\right|^{2} q = n(z,t).
\end{equation}

The ASE noise $\boldsymbol{n}(z,t)=[n_1(z,t), \,n_2(z,t)]$ in \eqref{Manakov.au} is a normalised version of $N(\ell,\tau)$, and is assumed to have the following correlation properties
\begin{equation}\label{noise}
\begin{split}
&\mathbb {E}  \left[ n_{i}(z,t) \right] = \mathbb{E} \left[ n_{i}(z,t) \,n_{j}(z',t') \right] = 0,
\\
&\mathbb {E} \left[ n_{i}(z,t) \, \bar{n}_{j}(z',t') \right] =
D \, \delta_{ij} \, \delta\left(z-z'\right)\delta\left(t-t'\right),
\end{split}
\end{equation}
with $i,j\in\{1,2\}$, with $\delta_{ij}$ being a Kronecker symbol, $\mathbb {E} \left[\cdot\right]$ is the mathematical expectation operator, and $\delta\left(\cdot\right)$ is the Dirac delta function. The correlation properties \eqref{noise} mean that each noise component $n_{i}(z,t)$ is assumed to be a zero-mean, independent, white circular Gaussian noise. The scalar case follows by considering a single noise component only.

The noise intensity $D$ in \eqref{noise} is (in dimensionless units)
\begin{equation}\label{D}
D = \sigma_{0}^2 \frac{\ell_0}{P_0 \tau_0}=\frac{\sigma_0^2}{\sqrt{\gamma^{*} \left|\beta_2\right| P_0^3}},
\end{equation}
where $\sigma_{0}^2$ is the spectral density of the noise, with real world units $\left[\mathrm{W}/\left(\mathrm{km} \cdot \mathrm{Hz}\right)\right]$. For ideal DRA, this $\sigma_{0}^2$ can be expressed through the optical fibre and transmission system parameters as follows: $\sigma_{0}^2 =  \alpha_\textnormal{fibre}  K_{T} \cdot h\nu_{0}$, where $\alpha_\textnormal{fibre}$ is the fibre attenuation coefficient, $h\nu_{0}$ is the average photon energy, $K_{T}$ is a temperature-dependent phonon occupancy factor \cite{ekw10}.

\begin{figure*}[tpb]
\begin{center}
\includegraphics[width=0.75\textwidth]{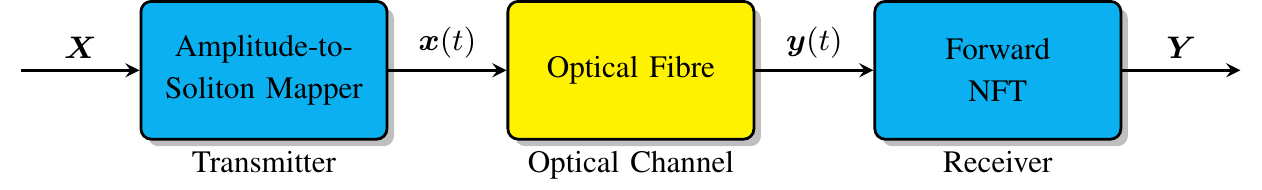}
\end{center}
\caption{System model under consideration. The symbols $\boldsymbol{X}=[X_{1},X_{2},X_{3},...]$ are converted to amplitudes, and then mapped to a waveform $\boldsymbol{x}(t)$. The noisy received waveform $\ybt$ is obtained by propagating $\xbt$ in \eqref{Manakov.au}. The forward NFT processes the waveform $\ybt$ symbol-by-symbol, and gives a soft estimate of the transmitted symbols $\boldsymbol{Y}=[Y_{1},Y_{2},Y_{3},...]$.}
 \label{Fig:CT}
\end{figure*}

From now on, all the quantities in this paper are in normalised units unless specified otherwise. Furthermore, we define the continuous time channel as the one defined by the normalised ME and the NSE. This is shown schematically in the inner part of Fig.~\ref{Fig:CT}, where the transmitted and received waveforms are $\xbt\equiv\q(0,t)$ and $\ybt\equiv\q(Z,t)$, respectively, where $Z$ is the propagation distance.

\subsection{Fundamental Soliton Solutions}\label{Sec:CT.Solitons}

It is known that the noiseless $\left(\n(z,t)=\boldsymbol{0}\right)$ ME \eqref{Manakov.au} possesses a special class of solutions, the so-called fundamental bright solitons.\footnote{Fundamental solitons are ``bright'' only for the focusing case we consider in this paper, i.e., for anomalous dispersion.} In general, the Manakov fundamental soliton is fully characterised by 6 parameters \cite{dpy06} (4 in the NSE case): frequency (also having the meaning of velocity in some physical applications), phase, phase mismatch, centre-of-mass position, polarisation angle, and amplitude (the latter is inversely proportional to the width of the soliton). In this paper we consider amplitude-modulated solitons, and thus, no information is carried by the other 5 parameters. The initial values of these 5 parameters can therefore be set to arbitrary values. In this paper, all of them have been set to zero. For the initial frequency, this can be further motivated to avoid deterministic pulse walk-offs. As for the initial phase, phase mismatch, and centre-of-mass position, as we shall see in the next section, their initial values do not affect the marginal amplitude channel law. Under these assumptions, the soliton solution at $z=0$ is given by \cite{lk98,dpy06}
\begin{equation}\label{soliton.1}
\q(0,t) = [q_1(0,t),q_2(0,t)]=\left[ \cos\beta_0, \,  \sin \beta_0 \right] A\sech (A t),
\end{equation}
where $A$ is the soliton amplitude and $0<\beta_0<\pi/2$ is the polarisation angle. The value of $\beta_0$ can be used to control how the signal power is split across the two polarisations.

For any $\beta_0$, the Manakov soliton solution after propagation over a distance $Z$ with the initial condition given by (\ref{soliton.1}), is expressed as
\begin{align}\label{soliton.1.5}
\q(Z,t) &= \left[  \cos \beta_0, \,  \sin \beta_0 \right] A \sech (A t) \exp \left( \frac{\imath A^{2} Z}{2} \right)\\
\label{soliton.2}
	      &= \q(0,t) \exp \left( \frac{\imath A^{2} Z}{2} \right).
\end{align}
The soliton solution for the NSE in \eqref{NLSE.au} can be obtained by using $\beta_0=0$ in \eqref{soliton.1}--\eqref{soliton.2}\footnote{This corresponds to the case where all the signal power is transmitted in the first polarisation.}, which gives
\begin{equation}\label{soliton.1.NLSE}
q(0,t) = A \sech \left(A t\right),
\end{equation}
and
\begin{align}\label{soliton.2.NLSE}
\nonumber
q(Z,t) & = A \sech \left(A t\right) \exp \left( \frac{\imath A^{2} Z}{2} \right)\\
	& = q(0,t) \exp \left( \frac{\imath A^{2} Z}{2} \right).
\end{align}

As shown by \eqref{soliton.2} and \eqref{soliton.2.NLSE}, the solitons in \eqref{soliton.1} and \eqref{soliton.1.NLSE} only acquire a phase rotation after propagation. When the noise is not zero, however, these solutions will change. This will be discussed in detail in the following section.

\section{Discrete-time Channel Model}\label{Sec:DT}

\subsection{Amplitude-modulated Solitons: One and Two Polarisations}\label{Sec:DT.4and6}

\begin{figure*}[tpb]
\begin{center}
\includegraphics[width=\textwidth]{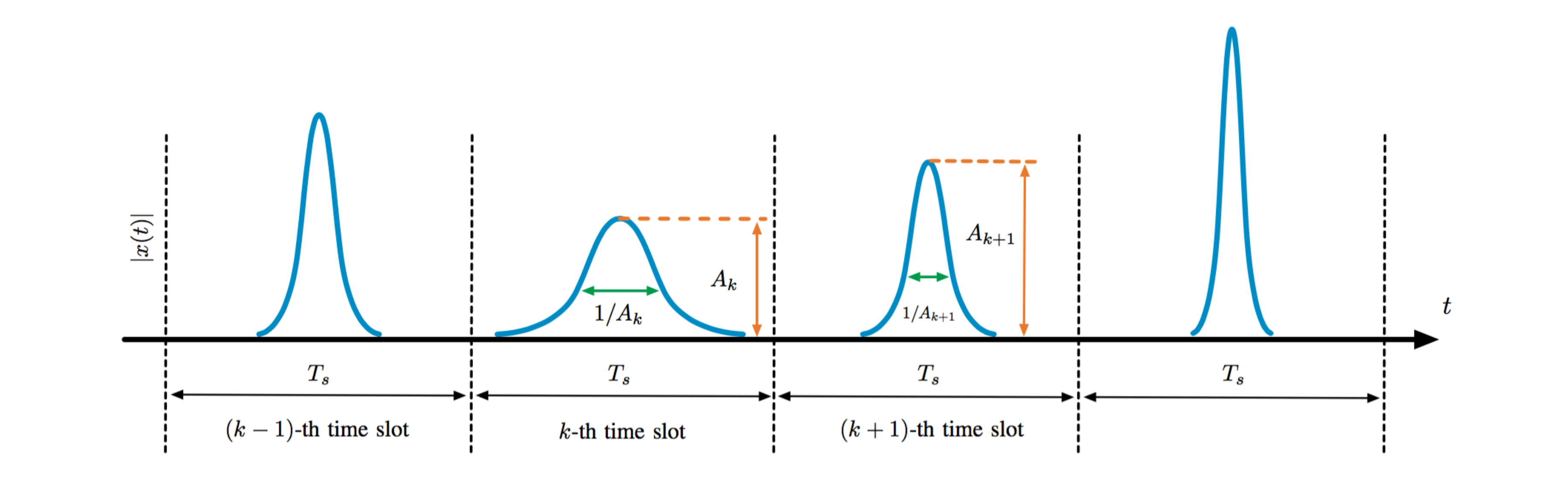}
\end{center}
\caption{\label{fig:sol-seq} Schematic visualisation of the amplitude modulation of soliton sequence (scalar NSE case).}
\end{figure*}

We consider a continuous-time input signal $\xbt=[x_1(t), \, x_2(t)]$ of the form
\begin{equation}\label{signal.1}
\xbt = \sum_{k=1}^\infty \s_k(t),
\end{equation}
where $\s_k(t)=[s_{k,1}(t), \, s_{k,2}(t)]$ and $k$ is the discrete-time index. Motivated by the results in Sec.~\ref{Sec:CT.Solitons}, the pulses $\s_k(t)$ are chosen to be
\begin{equation}\label{signal.2}
\s_{k}(t) = \left[ \cos\beta_0, \, \sin \beta_0 \right] \, A_{k} \sech\left[A_k(t - k T_s)\right],
\end{equation}
where $T_s$ is the symbol period. In principle, it is also possible to encode information by changing the polarisation angle $\beta_0$ from slot to slot. However, in this paper, we fix its value to be the same for all the time slots corresponding to a fixed (generally elliptic) degree of polarisation. Thus, the transmitted waveform corresponds to soliton amplitude modulation, which is schematically shown in Fig.~\ref{fig:sol-seq} for the scalar (NSE) case.

At the transmitter, we assume that symbols $X_{k}$ are mapped to soliton amplitudes $A_{k}$ via $A_{k}=X_{k}^2$. This normalisation is introduced only to simplify the analytical derivations in this paper. To avoid soliton-to-soliton interactions, we also assume that the separation $T_s$ is large, i.e., $\exp(-A_{k} T_s) \ll 1$, $\forall k$. The receiver in Fig.~\ref{Fig:CT} is assumed to process the received waveform during a window of $T_s$ via the forward NFT \cite{m74,mm15} and returns the amplitude of the received soliton, which we denoted by $R_{k}=Y_{k}^2$.

Before proceeding further, it is important to discuss the role of the amplitudes $A_k$ on a potential enhancement of soliton-soliton interactions. The interaction force prefactor is known to scale as the amplitude cubed \cite[Chapter 9.2]{hk95}, \cite[Chapter 5.4]{Iannone}. However, the interaction also decays exponentially as $\exp(-A_k T_s)$. This exponential decay dominates the interaction, and thus, considering very large amplitudes (or equivalently, very large powers, as we will do later in the paper), is in principle not a problem. At extremely large amplitudes, however, the model used in this paper is invalid for different reasons: higher order nonlinearities should be taken into account. This includes stimulated Brilloin scattering (for very large powers) or Raman scattering (for very short pulses). Studying these effects is, however, out of the scope of this paper.

We would also like to emphasise that for a fixed pulse separation $T_s$, the channel model we consider in this paper is not applicable for low soliton amplitudes. This is due to two reasons. The first one is that for low amplitude solitons, the perturbation theory used to derive the channel law becomes inapplicable as the signal becomes of the same order as noise. Secondly, low amplitude solitons are also very broad, and thus, nonnegligible soliton interactions are generated. These two cases can be overcome if the soliton amplitudes are always forced to be larger than certain cutoff amplitude $\coa$, which we will now estimate. For the first case (noise-limited), the threshold $\coaN$ is proportional to $\sigma_N^2$. In the second case (interaction-limited), the threshold is proportional to the symbol rate, i.e., $\coaS \propto T_s^{-1}$. This shows that for fixed system parameters, the threshold $\coa =\max\{\coaN,\coaS\}$ is a constant. The implications of this will be discussed at the end of Sec.~IV.

Having defined the transmitter and receiver, we can now define a discrete-time channel model, which encompasses the transmitter, the optical fibre, and the receiver, as shown in Fig.~\ref{Fig:CT}. Due to the assumption on solitons well-separated in time, we model the channel as memoryless, and thus, from now on we drop the time index $k$. This memoryless assumption is supported by additional numerical simulations we performed, which are included in Appendix \ref{sec:memoryless}. Nevertheless, at this point it is important to consider the implications of a potential mismatch between the memoryless assumption of the model and the true channel in the context of channel capacity lower bounds. In particular, if in some regimes (e.g., low power or large transmission distances) the memoryless assumption would not hold, considering a memoryless channel model would result in \emph{approximated} lower bounds on the channel capacity. Provable lower bounds can be obtained by using mismatched decoding theory \cite{alvkz06} (as done in \cite[Sec. III-A and III-B]{sar2016}) or by considering an average memoryless channel (as done in \cite[Sec. III-F]{ekw10}). Although both approaches can in principle be used in the context of amplitude-modulated solitons, they both rely on having access to samples from the \emph{true channel}, and not from a (potentially memoryless) model. Such samples can only be obtained through numerical simulations or an optical experiment, which is beyond the scope of this paper. In this context, the channel capacity lower bounds in Sec.~\ref{Sec:Main}, should be considered as a first step towards more involved analyses.

The conditional probability density function (PDF) for the received soliton amplitude $R$ given the transmitted amplitude $A$ was obtained in \cite[eq.~(15)]{dpy06} using standard perturbative approach and the Fokker-Planck equation method. The result can be expressed as a noncentral chi-squared distribution
\begin{equation}
p_{R|A} (r|a) =
\frac{1}{\sigma_{N}^2} \, \frac{r}{a} \, \exp \left( - \frac{a+r}{\sigma_{N}^2} \right)
I_{2} \left( \frac {2 \sqrt{a r}}{\sigma_{N}^2}\right),
\label{PDF.n3}
\end{equation}
where
\begin{equation}\label{sigmaN2}
\sigma_{N}^2 = D \cdot \frac{Z}{2}
\end{equation}
is the normalised variance of accumulated ASE noise, and $I_2(\cdot)$ is the modified Bessel function of the first kind of order two. The expression in \eqref{PDF.n3} is a noncentral chi-squared distribution with six degrees of freedom (see, e.g., \cite[eq.~(29.4)]{Johnson}) providing non-Gaussian statistics for Manakov soliton amplitudes. By making the change of variables $Y=\sqrt{R}$, and using $X=\sqrt{A}$, the PDF in \eqref{PDF.n3} can be expressed as
\begin{equation}
p_{Y|X}(y|x) =
\frac{2}{\sigma_{N}^2} \, \frac{y^3}{x^2}  \exp \left( - \frac{x^2+y^2}{\sigma_{N}^2} \right)
I_{2} \left( \frac {2 xy}{\sigma_{N}^{2}}\right),
\label{originalPDF.n3}
\end{equation}
which corresponds to the noncentral chi-distribution with six degrees of freedom. An extra factor $2y$ before the exponential function comes from the Jacobian.

For the NSE, it is possible to show that the channel law becomes \cite{spd15,dty03,dty05}
\begin{equation}
p_{Y|X}(y|x) = \frac{2}{\sigma_{N}^2} \, \frac{y^2}{x}  \exp \left( - \frac{x^2+y^2}{\sigma_{N}^2} \right)
I_{1} \left( \frac {2 xy}{\sigma_{N}^{2}}\right),
\label{originalPDF.n2}
\end{equation}
which corresponds to a noncentral chi-distribution with four degrees of freedom.

We note that although in this paper we only consider an amplitude modulation $A_k$ (or in the NFT terms the imaginary part of each discrete eigenvalue), it is possible to include other discrete degrees of freedom corresponding to various soliton parameters in \eqref{signal.2} in order to improve the achievable information rates. This is, however, beyond the scope of this paper. Furthermore, the channel models presented in this section were obtained via a perturbative treatment, and thus, in the context of soliton/eigenvalue communications they are technically valid only at high SNR.\footnote{More precisely, when the total soliton energy in the time slot is much greater than that of the ASE noise.} Despite that, in the current paper we will also study capacity lower bounds of a general noncentral chi-channel with arbitrary number of degrees of freedom any range of SNR. While admittedly the low-SNR region is currently only of interest when $n=1$ (noncoherent phase channel) we believe its generalization for $n>1$ can still be of interest for the new generation of nonlinear optical regeneration systems

\subsection{Generalised Discrete-time Channel Model}\label{Sec:DT.2n}

The results in the previous section show that both scalar and vector soliton channels can be modelled using the same class of the noncentral chi-distribution with an even number of degrees of freedom $2n$, with $n=2,3$. The simplest channel of this type corresponds to $n=1$, which describes a fibre optical communication channel with zero-dispersion\cite{tdy03} as well as the noncoherent phase channel studied in \cite{KatzShamai} (see also \cite{Lapidoth}). Motivated by this, here we consider a general communication channel described by the noncentral chi-distribution with an arbitrary (even) degrees of freedom $2n$. Although we are currently not aware of any physically-relevant communication system that can be modelled with $n\geq 4$, we present results for arbitrary $n$ to provide an exhaustive treatment for channels of this type.

\begin{figure}[tpb]
\begin{center}
\includegraphics[width=\columnwidth]{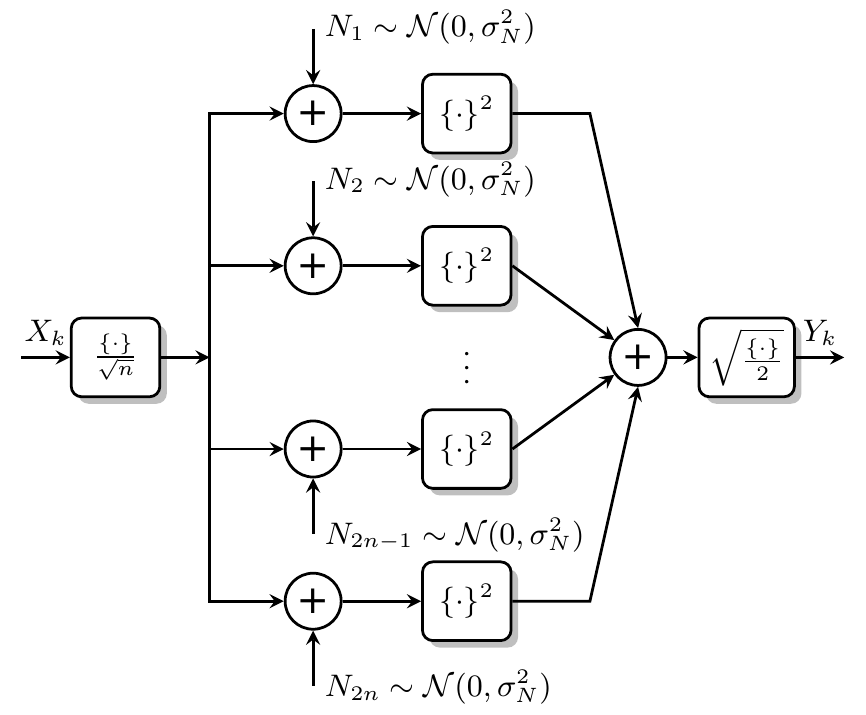}
\end{center}
\caption{Generalised discrete-time channel model: noncentral chi-channel with $2n$ degrees of freedom.}
\label{Fig:DT}
\end{figure}

The channel in question is therefore modelled via the PDF corresponding to noncentral chi-distribution
\begin{equation}
p_{Y|X}(y|x) =
\frac{2}{\sigma_{N}^2} \, \frac{y^{n}}{x^{n-1}}  \exp \left( - \frac{x^2+y^2}{\sigma_{N}^2} \right)
I_{n-1} \left( \frac {2 xy}{\sigma_{N}^2}\right),
\label{channel-law}
\end{equation}
with $n\in\mathbb{N}$ and where $\mathbb{N} \triangleq \left\{1,2,3,...\right\}$. This channel law corresponds to the following input-output relation
\begin{eqnarray}\label{channel_model}
Y^2 = \frac{1}{2 } \sum_{i=1}^{2n} \, \left(\frac{X}{\sqrt n} +  N_{i} \right)^2,
\end{eqnarray}
where $\left\{N_i\right\}_{i=1}^{2n}$ is a set of independent and identically distributed Gaussian random variables with zero mean and variance $\sigma_{N}^{2}$. The above input-output relationship is schematically shown in Fig.~\ref{Fig:DT}, which particularises to \eqref{originalPDF.n3} and \eqref{originalPDF.n2}, for $n=3$ and $n=2$, respectively.

\section{Main Results}\label{Sec:Main}

In this section, we study capacity lower bounds of the channel in \eqref{channel-law}. We will show results as a function of the effective SNR defined as $\rho \triangleq \sigma_{S}^2/\sigma_{N}^2$, where $\sigma_{S}^2$ is the second moment of the input distribution $p_X$ and $\sigma_N^2$ is given by \eqref{sigmaN2}. The value of $\sigma_{S}^2$ also corresponds to the average soliton amplitude, i.e., $\sigma_{S}^{2}= \mathbb {E} \left[ X^{2} \right]=\mathbb {E}\left[ A \right]$. It can be shown that for given system parameters, the noise power (in real world units) is constant and proportional to $\sigma_N^2$, and the signal power (in real world units) is proportional to $\sigma_{S}^2$. The parameter $\rho$ therefore indeed corresponds to an effective SNR.

As previously explained, the inter-symbol interference due to pulse interaction can be neglected due to the large enough soliton separation assumed, and thus, the channel can be treated as a \emph{memoryless} (see Appendix~\ref{sec:memoryless} for more details). The channel capacity, in bits per channel use, is then given by\cite{Shannon,CoverThomas}
\begin{equation}
\label{Capacity_def}
C(\rho) \triangleq \max_{p_{X}(x): \,\, \mathbb{E} \left[X^{2}\right] \leq \sigma_{S}^{2}} \, \MI(\rho),
\end{equation}
where
\begin{align}\label{MI_def.0}
\MI(\rho) &\triangleq\mathbb{E} \left[\log_2\frac{p_{X,Y}(X,Y)}{p_{X}(X) \cdot p_{Y}(Y)}\right] \\
\label{MI_def}
&=h_{Y}(\rho) - h_{Y|X}(\rho),
\end{align}
and where $h_Y(\rho) \triangleq -\,\mathbb{E} \left[\log_2 p_Y(Y) \right]$ and $h_{Y|X}(\rho) \triangleq -\, \mathbb{E} \left[\log_2 p_{Y|X}(Y|X) \right]$ are the output and conditional differential entropies, respectively. The optimisation in \eqref{Capacity_def} is performed over all possible statistical distributions $p_{X}(x)$ that satisfy the power constraint. In our case this constraint corresponds to a fixed second moment of the input symbol distribution or, equivalently, to a fixed average signal power in a given symbol period.

The exact solution for the power-constrained optimisation problem \eqref{Capacity_def} with the channel law \eqref{channel-law} is unknown. For the noncentral chi-distribution with 2 degrees of freedom (i.e., to the noncoherent additive noise channel), it was shown \cite{KatzShamai} that the capacity-achieving distribution is discrete with an infinite number of mass points. To the best of our knowledge, that proof has not been extended to higher number of degrees of freedom, however, we expect that will be the case for \eqref{channel-law} too.

In this paper, we do not aim at finding the capacity-achieving distribution, but instead, we study lower bounds on the capacity. We do this because the capacity problem is in general very difficult, but also because of the  relevance of having nondecreasing lower bounds on the capacity for the optical community. To obtain a lower bound on the capacity, we will simply choose an input distribution $p_{X}(x)$ (as done in, e.g., \cite{ms01,spd15}). Without claiming the generality, we, however, consider four important candidates for the input distribution.
First, following \cite{spd15}, we use symbols drawn from a Rayleigh distribution
\begin{equation}\label{Rayleigh}
p_{X}(x) = \frac{2x}{\sigma_{S}^{2}} \exp \left( - \frac{x^2}{\sigma_{S}^{2}} \right), \quad x\in {[0,\infty)}.
\end{equation}
As we will see later, this input distribution is not the one giving the highest lower bound. However, it has one important advantage: it allows some \emph{analytical} results for the mutual information. The other three distributions are considered later in this section as numerical examples.

The next two Lemmas provide an exact closed-form expression for the conditional differential entropy $h_{Y|X}(\rho)$ and an asymptotic expression for the output differential entropy $h_{Y}(\rho)$.

\begin{lemma}\label{Lemma_CE}
For the channel in \eqref{channel-law} and the input distribution \eqref{Rayleigh}
\begin{align}\label{cond-entrop}
\nonumber
h_{Y|X}(\rho)=& \left(2\rho+n-\frac{n}{2}\psi(n)\right)\log_2 e - 1 \\
&+\frac{n-1}{2} \left(\log_2 \rho + \psi(1)\log_2 e \right) \nonumber \\
&- \frac{n \log_2 e}{2}\frac{\rho}{\rho+1}\, \Phi\!\left(\frac{\rho}{\rho+1},1,n\right) \nonumber \\
&-  \rho^{-1}\left (\frac{\rho+1}{\rho} \right)^{(n-1)/2} F_n(\rho) \log_2 e ,
\end{align}
where $\psi(x) \triangleq d \natlog \Gamma(x)/dx$ is the digamma function and $\Phi(\alphaproof,1,n)$ is the special case of the Lerch transcendent function \cite[eq.~(9.551)]{GR}
\begin{equation}\label{PHI}
\Phi(\alphaproof,1,n) \triangleq -\frac{\natlog(1-\alphaproof)}{\alphaproof^n} -\sum_{k=0}^{n-2} \frac{\alphaproof^{k+1-n}}{k+1}.
\end{equation}
The function $F_n(\rho)$ is defined as
\begin{equation}\label{F2}
F_{n}(\rho) \triangleq \intop_0^\infty \xi K_{n-1}(\sqrt{1+\rho^{-1}}\,\xi) \, I_{n-1}(\xi) \natlog \left[I_{n-1}(\xi)\right] \, d\xi,
\end{equation}
and $K_n(x)$ is the modified Bessel function of the second kind of order $n$.
\end{lemma}
\begin{IEEEproof}
See Appendix~\ref{Lemma.App}.
\end{IEEEproof}

\begin{lemma}\label{Lemma_HY}
For the channel in \eqref{channel-law} and the input distribution \eqref{Rayleigh}
\begin{align}
\nonumber
h_{Y}(\rho)=\frac{1}{2}\log_2 \rho +  \left(1 -\frac{\psi(1)}{2}\right) \log_2 e -1 + &O \left[\rho^{-1}\right],\\
\label{Lemma.lb}
		 &\rho \to \infty
\end{align}
\end{lemma}
\begin{IEEEproof}
See Appendix~\ref{sec:asymp}.
\end{IEEEproof}

The next theorem is one of the main results of this paper.
\begin{theorem}
\label{theorem1}
The MI for the channel in \eqref{channel-law} and the input distribution \eqref{Rayleigh} admits the following asymptotic expansion
\begin{align}\label{theorem1.eq}
\MI(\rho)& = \frac{1}{2}\log_2 \frac{e^{1 -\psi(1)}}{4\pi}\rho + O \left[\rho^{-1}\right], \quad \rho \to \infty.
\end{align}
\end{theorem}
\begin{IEEEproof}
We expand the function $F_n(\rho)$ in \eqref{F2} defining the conditional entropy in Lemma~\ref{Lemma_CE}. At fixed large $\rho$ the integrand  asymptotically decays as $\exp\left(-\xi/2\rho\right)$, i.e., with small decrement (which can be proven by a standard large argument asymptotes of the Bessel functions). This means that the main contribution to the integral comes from the asymptotic region $1 \lesssim \xi \lesssim \rho$ in most part of which the large argument expansion of both Bessel functions is indeed justified. Using it uniformly we obtain
\[
F_{n}(\rho)= 2 \rho ^2  + \frac{\rho}{2} \left[\log \frac{1}{\rho}  + 1 -\natlog 4\pi -\psi(1)  \right] + O \left[1\right],
\]
which used in \eqref{cond-entrop} gives the asymptotic expression
\begin{equation}
h_{Y|X}(\rho) = \frac{1}{2} \log_2 \pi e + O \left[\rho^{-1}\right], \quad \rho \to \infty.
\label{Hyx-asymp}
\end{equation}
The proof is completed by combining \eqref{Hyx-asymp} and \eqref{Lemma.lb} with \eqref{MI_def}.
\end{IEEEproof}

\begin{figure}[t!]
\begin{center}
\includegraphics[width=\columnwidth]{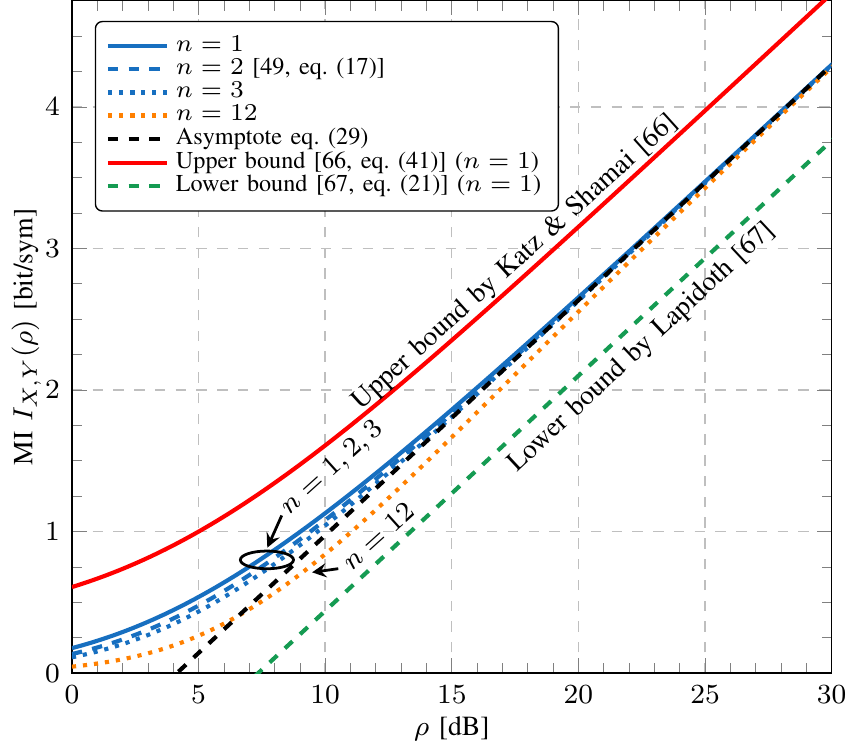}
\end{center}
\caption{The MI $\MI(\rho)$ in \eqref{MI_def} (numerically calculated) for the chi-distribution with different degrees of freedom and the channel model \eqref{channel-law}. The asymptotic estimate given by Theorem \ref{theorem1} is also shown. Lower and upper~bounds for $n=1$ are also shown.}
\label{Figure_3}
\end{figure}

The result in Theorem~\ref{theorem1} is a universal and $n$-independent expression. The expression in \eqref{theorem1.eq} shows that the capacity lower bound is asymptotically equivalent to half of logarithm of SNR plus a constant which is order-independent. Fig.~\ref{Figure_3} shows the numerical evaluation of $\MI (\rho)$ for $n=1,2,3,12$ obtained by numerically evaluating all the integrals in the exact expressions for the conditional and output entropies in \eqref{cond-entrop} and \eqref{E0}, as well as the asymptotic expression in Theorem~\ref{theorem1}. Interestingly, we can see that even in the medium-SNR region, the influence of the number of degrees of freedom on the MI is minimal, and the curves are quite close to each other. In this figure, we also include the lower and upper~bounds for $n=1$ given by \cite[eq.~(21)]{Lapidoth}  and \cite[eq.~(41)]{KatzShamai}, resp. These results show that the asymptotic results in Theorem~\ref{theorem1} correctly follow these two bounds.

\begin{figure}[t!]
\begin{center}
\includegraphics[width=\columnwidth]{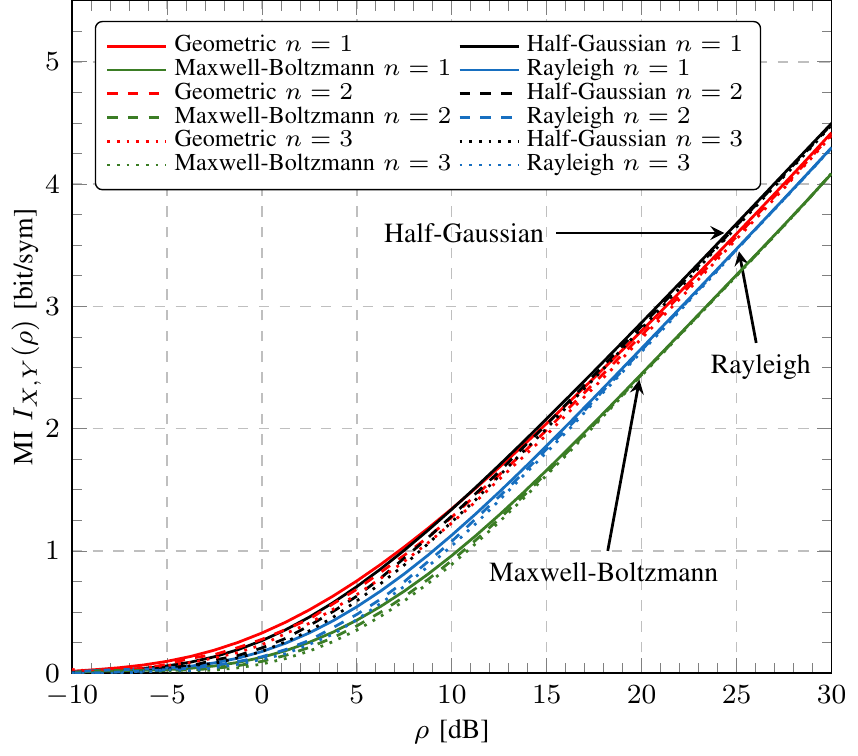}
\end{center}
\caption{MI estimates (by numerically evaluating \eqref{MI_def} via Monte-Carlo integration) for different trial continuous input distributions and different values of $n$ (different line types). Different distributions are shown with different colours.}
\label{MI_MC_contin}
\end{figure}

The main reason for considering a Rayleigh input distribution was that it yields a semi-analytical lower bound on the the capacity. In the following example, we consider three other input distributions and numerically calculate the resulting MI.
\begin{example}
Consider the geometric (exponential), half-Gaussian, and Maxwell-Boltzmann distributions given by
\begin{equation}\label{Geom}
p_{X}(x) = \frac{\sqrt{2}}{\sigma_{S}} \exp \left(-\frac{\sqrt{2} \, x}{\sigma_{S}}\right), \quad x\in {[0,\infty)},
\end{equation}
\begin{equation}\label{halfGauss}
p_{X}(x) = \frac{\sqrt{2}}{\sqrt{\pi} \sigma_{S}} \exp \left(-\frac{x^{2}}{2 \sigma_{S}^{2}}\right), \quad x\in {[0,\infty)},
\end{equation}
and
\begin{equation}\label{MaxBoltz}
p_{X}(x) = \frac{3 \sqrt{6} \, x^{2}}{\sqrt{\pi}\sigma_{S}^{3}} \exp \left(-\frac{3 \, x^{2}}{2 \sigma_{S}^{2}}\right), \quad x\in {[0,\infty)},
\end{equation}
respectively. The MIs for these three distributions for $n=1,2,3$ are shown in Fig.~\ref{MI_MC_contin} and show that the lower bound given by the geometric input distribution in \eqref{Geom} displays high MI in the low SNR regime ($\rho<10$~dB), whereas the half-Gaussian input distribution in \eqref{halfGauss} is better for medium and large SNR. On the other hand, the Maxwell-Boltzmann distribution in \eqref{MaxBoltz} gives the lowest MI for all SNR. Numerical results also indicate that all the presented MIs asymptotically exhibit an equivalent growth irrespective of the number of the degrees of freedom $2n$.

\end{example}

The following example considers the use of discrete constellations. In particular, we assume that the soliton amplitudes take values on a set $\mathcal{X}\triangleq \{x_{1},...,x_{M}\}$, where $M\triangleq \left|\mathcal{X}\right|=2^{m}$ is the cardinality of the constellation, and $m$ is a number of bits per symbol. The MI \eqref{MI_def} in this case can be evaluated as
\begin{align}\label{mi_ask}
\nonumber
\MI(\rho)&= \frac{1}{M} \sum_{x\in\mathcal{X}}  \, \intop_0^\infty p_{Y|X}(y|x) \\
&\qquad\qquad \cdot \log_{2}\frac{p_{Y|X}(y|x)}{\frac{1}{M} \sum_{x'\in\mathcal{X}}  p_{Y|X}(y|x')} \, dy,
\end{align}
where we assumed the symbols are equally likely.

\begin{figure}[t!]
\begin{center}
\includegraphics[width=\columnwidth]{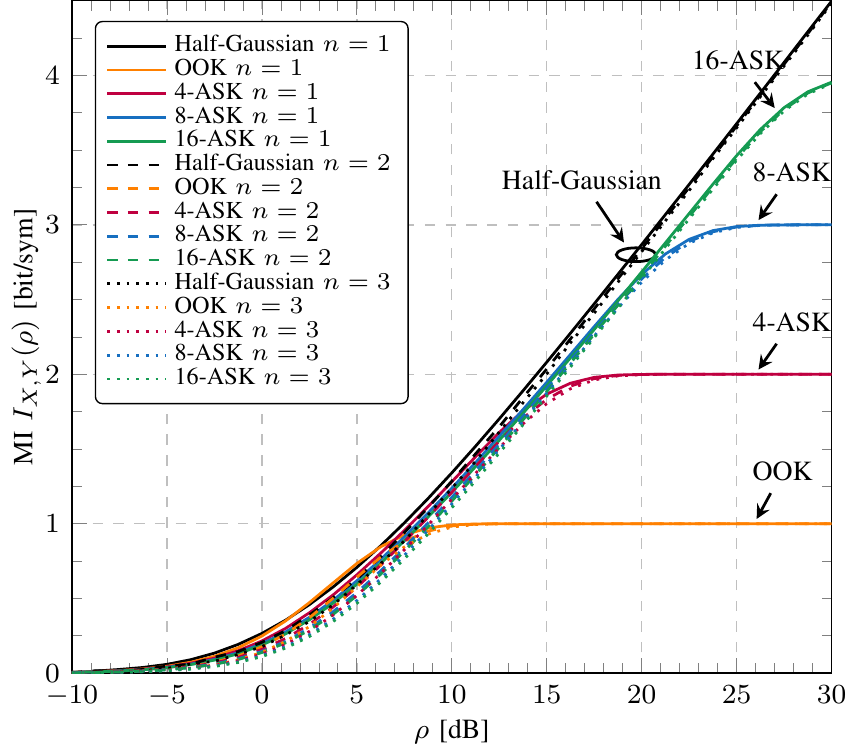}
\end{center}
\caption{MI estimates (numerically calculated) for equally-spaced $M$-ASK constellations with $M=\{2,4,8,16\}$ constellation points.}
\label{MI_MC}
\end{figure}

\begin{example}
Consider ASK constellations $\mathcal{X}=\{0,1,\ldots,M-1\}$ with $m=1,2,3,4$ and second moment $\sigma_{S}^{2}$, which correspond to OOK, 4-ASK, 8-ASK, and 16-ASK, respectively. The MI numerically evaluated for these constellations is shown in Fig.~\ref{MI_MC} for chi-channel with $n=1,2,3$. As a reference, in this figure we also show (black lines) the MI for the (continuous) half-Gaussian input distribution. The results in this figure show that in the low SNR regime, the use of  binary modulation is in fact better than the half-Gaussian distribution. This can, however, be remedied by using a geometric distribution, which, as shown in Fig.~\ref{MI_MC_contin}, outperforms the half-Gaussian distribution in the low SNR regime. In the high SNR regime, however, this is not the case.

\end{example}

Finally, let us address the impact of the cutoff $\coa$ we introduced in Sec.~III. All our results for continuous input distributions have been obtained for the input distributions that are not bounded away from zero (see (\ref{Rayleigh}), (\ref{Geom})--(\ref{MaxBoltz})). Therefore, symbols $X_k$ are generated below the threshold $\cox=\sqrt{\coa}$, where the channel law considered in this paper does not hold. We shall now only consider here the case of the Rayleigh input (\ref{Rayleigh}) as this distribution was used to obtain the main result of this section. We will prove that in the high-power (i.e., high SNR) regime, the effect of the cutoff on the achievable data rate tends to zero. To do so, we note that for fixed fibre parameters  and propagation distance, the cutoff $\cox^2=\coa=\max\{\coaN,\coaS\}$ is also fixed, while $\sigma_S^2=\rho \sigma_N^2$ grows linearly with SNR. In other words, one can achieve high SNR at the expense of high power solitons for fixed noise variance. One possible way of showing that the effect of the cutoff on the achievable rate is zero as SNR tends to infinity is to consider a transmitter which generates a dummy symbol every time $X_k\leq \cox$. The value of the threshold $\cox$ is message-independent and thus, can be assumed to be known to the receiver which will  discard sub-threshold symbols. This allows us to keep the main results of the paper at the expense of a data rate loss (since part of the time, dummy symbols are transmitted). The probability of such ``outage'' event $\eta$ is given by an the integral of the input distribution from zero to the threshold. For the Rayleigh input PDF (24) this probability is given by $\eta=1-\exp \left(-\coa/\sigma_S^2\right)$ (see \eqref{Xo.2}--\eqref{Xo.5}).  Therefore asymptotically $\eta(\rho)\approx \coa/(\rho \sigma_N^2) \to 0$ when $\rho \to \infty$. The average rate loss is then given by $1-\eta(\rho)$, which tends to zero as $\rho \to \infty$.

An alternative and more rigorous solution to the problem above is to consider directly the difference between the MI asymptote obtained in the current paper (i.e., Theorem 3) and that obtained by a truncated input Rayleigh distribution which simply does not generate sub-threshold symbols. This difference can be shown to tend to zero as $\rho\to \infty$. This proof is given in Appendix~\ref{sec:rateloss}.

\section{Conclusions}\label{Sec:Conclusions}

A non-Gaussian channel model for the conditional PDF of well-separated (in time) soliton amplitudes was used to study lower bounds on the channel capacity. Results for propagation of signals over a nonlinear optical fibre using one and two polarisations were presented. The results in this paper demonstrated both analytically and numerically that there exist lower bounds on the channel capacity that display an unbounded growth with the effective SNR, similarly to the linear Gaussian channel. All the results in this paper are given in bit per channel use only, and thus, they should be considered as a first step towards analysing the more practically relevant problem of  channel capacity in bit per second per unit bandwidth. This is a considerably more challenging  problem, which is left for further investigation.

Apart from the ME soliton channel model this paper also studied lower bounds on the capacity of an abstract general noncentral chi-channel with arbitrary number of degrees of freedom. Similar channel models appear in the study of relatively general systems of noise-driven coupled nonlinear oscillators \cite{pd05}. Therefore, we believe that the results for large number of degrees of freedom might also some day find applications in nonlinear communication channels.

The results obtained in this paper for the general noncentral chi-Channel are true capacity lower bounds for that channel model. For the case of the application considered in this paper (amplitude-modulated soliton systems), however, the  presented analysis was based on a perturbative-based model which holds at high SNR. This model also does not consider potential interaction between solitons, and thus, the results in this paper are limited to solitons well separated in time. Another way of interpreting these results is that the obtained expressions are  approximated lower bounds on the capacity of the true channel. Bounds that consider memory effects are left for further investigation. Furthermore, another interesting open research problem is the derivation of capacity upper bounds for amplitude-modulated soliton systems. This is also left for further investigation.

\appendices

\section{Memoryless property of the discrete-time channel model}\label{sec:memoryless}

In this section, we present numerical simulations to verify the memoryless assumption for the discrete channel model in Sec.~\ref{Sec:DT}. To this end, we simulated the propagation of sequences of $N=10$ soliton symbols through the scalar waveform channel given by \eqref{NLSE.au}. Two launch powers ($-1.5$ and $1.45$~dBm) and two propagation distances ($500$~km and $2000$~km) are considered. The simulations were carried out via the standard split-step Fourier method. The soliton amplitudes were generated as i.i.d. samples from a Rayleigh input distribution (see \eqref{Rayleigh}) and the variance of $X$ was chosen to be 1.25 and 20, so that the resulting soliton waveforms have powers of $-1.5$ and $1.45$~dBm, respectively. The transmitted waveform $x(\tau)$ was created using \eqref{signal.1} at a symbol rate of $1.7$~GBd. To guarantee an accurate simulation, the time-domain samples were taken every $4.6$~ps and the step size was $0.1$~km. White Gaussian noise was added at each step to model the ideal DRA process. The simulation parameters are similar to those used in \cite{yrft14} and are summarised in Table~\ref{system_tab}.

\begin{table}[!t]
\centering
\caption{Simulation system parameter}
\label{system_tab}
\begin{IEEEeqnarraybox}[
\IEEEeqnarraystrutmode
\IEEEeqnarraystrutsizeadd{2pt}{1pt}]{v/c/v/c/v}
\IEEEeqnarrayrulerow\\
&\mbox{Parameter}&&\mbox{Value}&\\
\IEEEeqnarraydblrulerow\\
\IEEEeqnarrayseprow[3pt]\\
&\mbox{Carrier frequency} \, (\nu_0) && 193.41\,\mathrm{THz}&\IEEEeqnarraystrutsize{0pt}{0pt}\\
\IEEEeqnarrayseprow[3pt]\\
\IEEEeqnarrayrulerow\\
\IEEEeqnarrayseprow[3pt]\\
&\mbox{Fibre attenuation} \, (\alpha_\textnormal{fibre}) && 0.20 \, \mathrm{dB} \,\,\mathrm{km}^{-1} &
\IEEEeqnarraystrutsize{0pt}{0pt}\\
\IEEEeqnarrayseprow[3pt]\\
\IEEEeqnarrayrulerow \\
\IEEEeqnarrayseprow[3pt]\\
&\mbox{Fibre group-velocity dispersion} \, (\beta_2) && - 21.67 \,\mathrm{ps}^{2}\, \mathrm{km}^{-1} &
\IEEEeqnarraystrutsize{0pt}{0pt}\\
\IEEEeqnarrayseprow[3pt]\\
\IEEEeqnarrayrulerow\\
\IEEEeqnarrayseprow[3pt]\\
&\mbox{Fibre nonlinearity} \, (\gamma) && 2.0 \, \mathrm{W}^{-1} \mathrm{km}^{-1} &
\IEEEeqnarraystrutsize{0pt}{0pt}\\
\IEEEeqnarrayseprow[3pt]\\
\IEEEeqnarrayrulerow\\
\IEEEeqnarrayseprow[3pt]\\
&\mbox{Phonon occupancy factor} \, (K_{T}) && 1.13 &
\IEEEeqnarraystrutsize{0pt}{0pt}\\
\IEEEeqnarrayseprow[3pt]\\
\IEEEeqnarrayrulerow\\
\IEEEeqnarrayseprow[3pt]\\
&\mbox{Propagation distance} && 500 \,\mathrm{km}&
\IEEEeqnarraystrutsize{0pt}{0pt}\\
\IEEEeqnarrayseprow[3pt]\\
\IEEEeqnarrayrulerow \\
\IEEEeqnarrayseprow[3pt]\\
&\mbox{Propagation step-size} && 0.1 \,\mathrm{km}&
\IEEEeqnarraystrutsize{0pt}{0pt}\\
\IEEEeqnarrayseprow[3pt]\\
\IEEEeqnarrayrulerow
\end{IEEEeqnarraybox}
\end{table}

\begin{figure*}[thpb]
\centering
{\footnotesize (a) Launch Power is $-1.5$~dBm ($\Var[X]=1.25$)}\\
\vspace{1ex}
\begin{subfigure}[b]{0.45\textwidth}
\begin{center}
\includegraphics[width=\columnwidth]{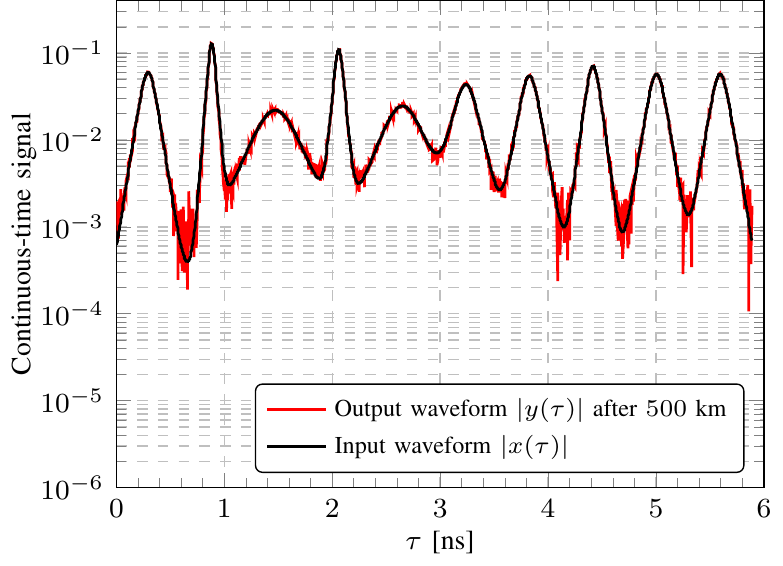}
\end{center}
\label{fig_7_a}
\end{subfigure}
\hfill
\begin{subfigure}[b]{0.45\textwidth}  
\centering 
\begin{center}
\includegraphics[width=\columnwidth]{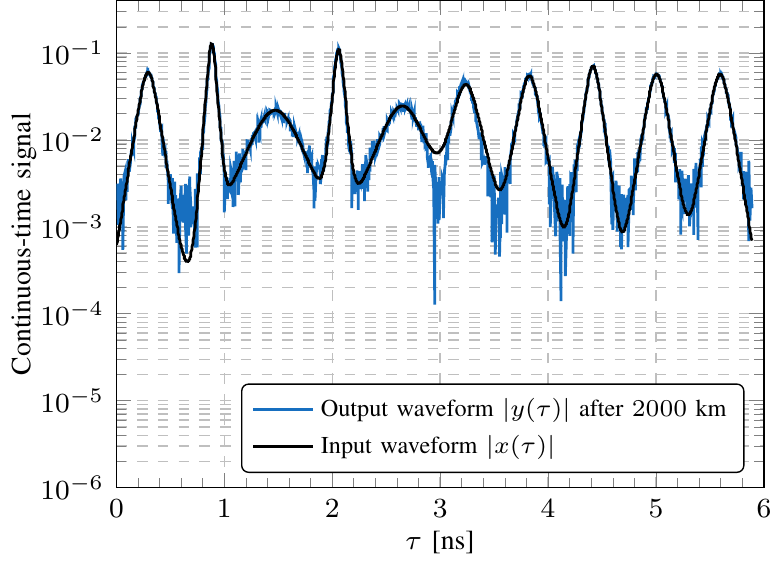}
\end{center}
\label{fig_7_b}
\end{subfigure}
\vfill 
\centering 
\vspace{-1ex}
{\footnotesize (b) Launch Power is $1.45$~dBm ($\Var[X]=20$)}\\
\vspace{1ex}
\begin{subfigure}[b]{0.45\textwidth}   
\begin{center}
\includegraphics[width=\columnwidth]{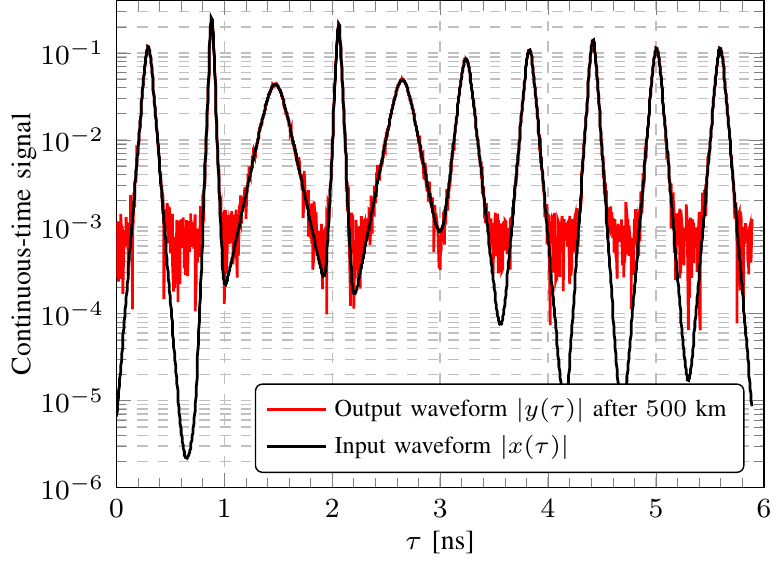}
\end{center}
\label{fig_7_c} 
\end{subfigure}
\hfill 
\begin{subfigure}[b]{0.45\textwidth}   
\begin{center}
\includegraphics[width=\columnwidth]{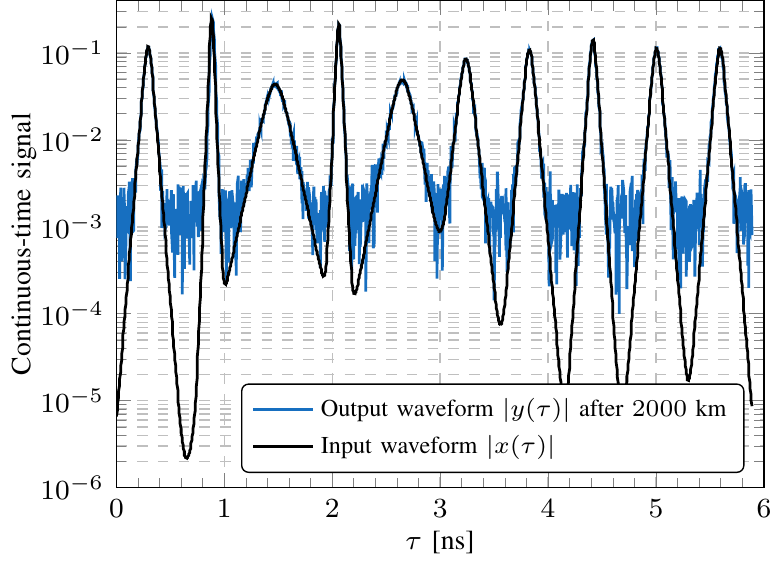}
\end{center}
\label{fig_7_d} 
\end{subfigure}
\caption{Continuous-time input $x(\tau)$ and output $y(\tau)$ soliton waveforms for 10 solitons and distributed noise due to IRA. Two launch powers are considered: (a) $-1.5$~dBm and (b) $1.45$~dBm. The solitons are propagated $500$ and $2000$~km.}
\label{Figure_7}
\end{figure*}

\begin{figure*}[thpb]
\centering
{\footnotesize (a) Launch Power is $-1.5$~dBm ($\Var[X]=1.25$)}\\
\vspace{1ex}
\begin{subfigure}[b]{0.475\textwidth}
\begin{center}
\includegraphics[width=0.9\columnwidth]{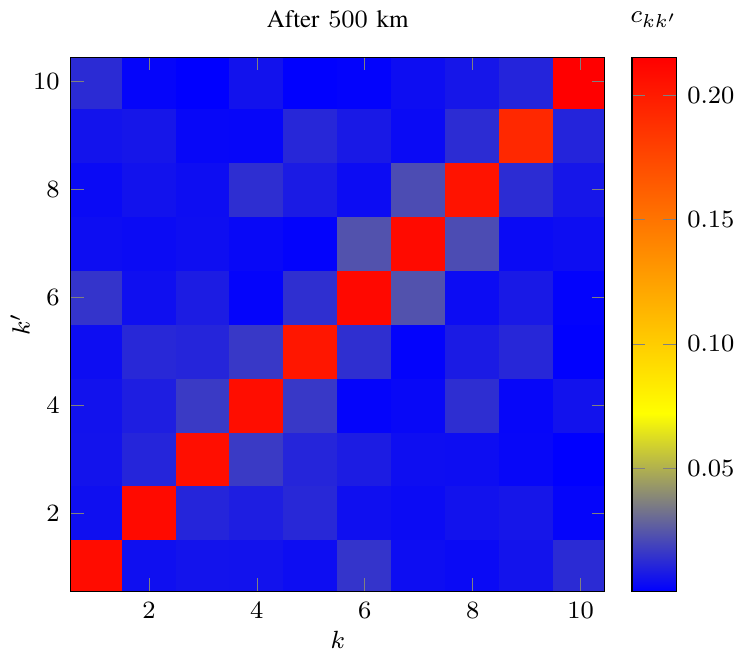}
\end{center}
\label{fig_8_a}
\end{subfigure} 
\hfill
\begin{subfigure}[b]{0.475\textwidth}  
\begin{center}
\includegraphics[width=0.9\columnwidth]{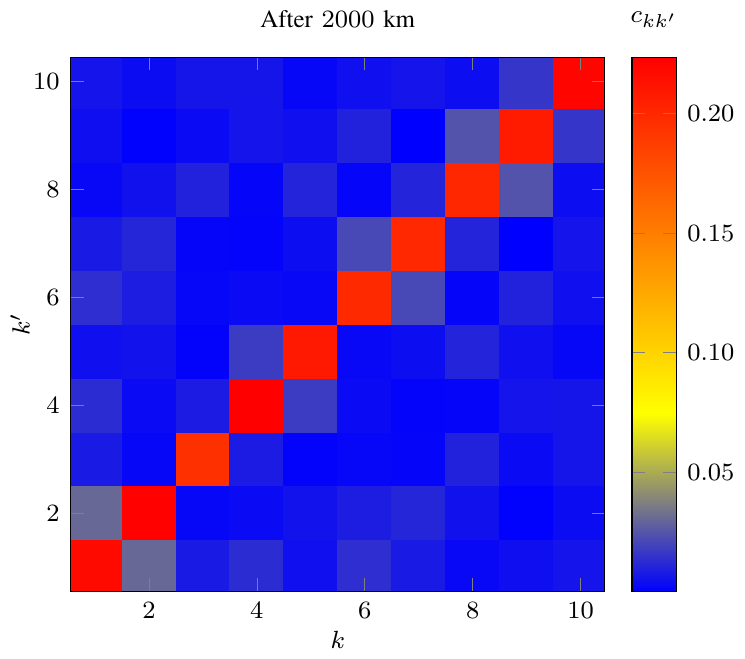}
\end{center}
\label{fig_8_b}
\end{subfigure}
\vfill 
\vspace{-2ex}
{\footnotesize (b) Launch Power is $1.45$~dBm ($\Var[X]=20$)}\\
\vspace{1ex}

\begin{subfigure}[b]{0.475\textwidth}   
\begin{center}
\includegraphics[width=0.9\columnwidth]{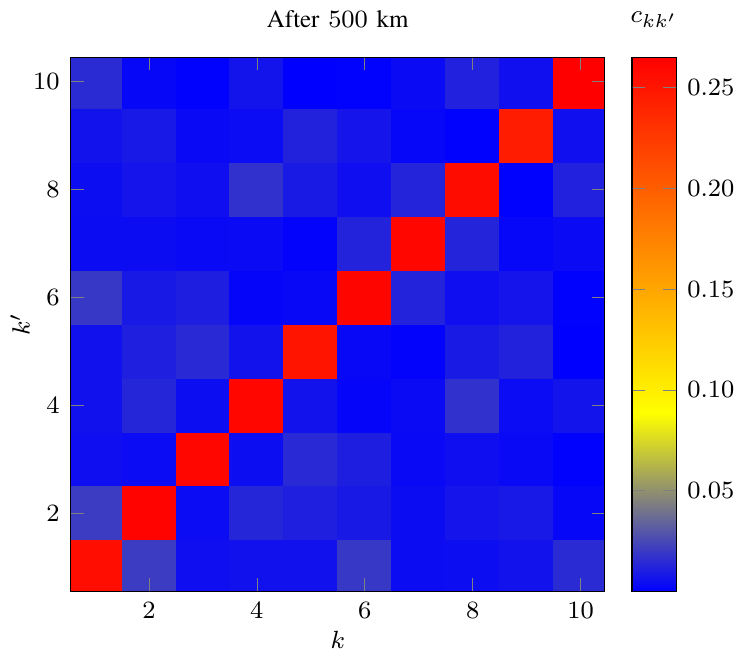}
\end{center}
\label{fig_8_c} 
\end{subfigure}
\hfill 
\centering
\begin{subfigure}[b]{0.475\textwidth}   
\begin{center}
\includegraphics[width=0.9\columnwidth]{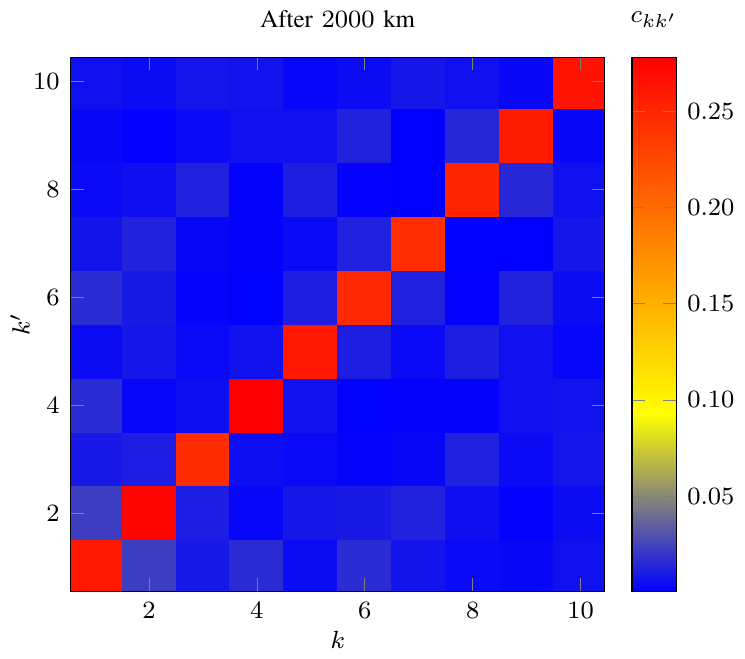}
\end{center}         
\label{fig_8_d}
\end{subfigure}
\caption{Normalised output symbol correlation matrices for the two launch powers and propagation distances in Fig.~\ref{Figure_7}.}
\label{corr-matrix}
\end{figure*}

Fig.~\ref{Figure_7} shows the waveforms before and after propagation through the channel given in \eqref{NLSE.au}. As expected, the received signal is a noisy version of the transmitted waveform, where the noise increases as the propagation distance increases. These results show that doubling the transmission distance and/or (approximately) doubling the launch power has very little effect in the soliton shapes.

The noisy waveforms shown in Fig.~\ref{Figure_7} were then used to obtain soliton amplitudes $\boldsymbol{Y} \triangleq \left[ Y_{1},Y_{2},\ldots,Y_{10} \right]$ via the forward NFT. Each amplitude is obtained by processing the corresponding symbol period via the spectral matrix method \cite[Sec.~IV-B]{yk14-2}. To test the memoryless assumption, we perform a simple correlation test. In particular, we consider the normalised output symbol correlation matrix, whose entries are defined as
\begin{equation}
c_{kk'} \triangleq \frac{\mathbb{E} \left[\left(Y_k-\mathbb{E}\left[Y_k\right]\right) \left(Y_{k'}-\mathbb {E}\left[Y_{k'}\right]\right)\right]}{\mathbb{E} \left[Y_k\right]\,\mathbb{E}\left[Y_{k'}\right]}.
\end{equation}
The obtained correlation matrices are shown in Fig.~\ref{corr-matrix}, where statistics were gathered by performing $10^3$ Monte-Carlo runs of the signal propagation. As we can see from Fig.~\ref{corr-matrix}, the matrices are almost diagonal. Since our communication channel is believed to be non-Gaussian, the absence of correlation does not of course necessarily imply the memoryless property (understood here as the statistical independence). However, it does constitute an important quantification of the qualitative criterion $\exp(-A_k T_s) \ll 1$ as given in Sec.~\ref{Sec:DT.4and6}.

\section{Proof of Lemma~\ref{Lemma_CE}}\label{Lemma.App}

The MI is invariant under a simultaneous linear re-scaling of the variables $x \to x/\sigma_N$ and $y \to y/\sigma_N$. For notation simplicity, and without loss of generality, throughout this proof we thus assume $\sigma_N^2=1$. Furthermore, we study the conditional entropy as a function of $\rho= \sigma_S^2$ and all the results will be given in nats.

We express the conditional differential entropy as
\begin{align}
\label{cond_entr1.0}
h_{Y|X}(\rho)	=& - \intop_{0}^\infty\intop_{0}^\infty p_{X,Y}(x,y) \natlog p_{Y|X}(y|x) \, dy \, dx\\	
\nonumber
				=& - \natlog 2 - n\,\mathbb{E}\left[\natlog Y\right] +\left(n-1\right)\mathbb{E}\left[\natlog X\right] \\
\label{cond_entr1}
				&+\mathbb{E}\left[X^2 \right]+\mathbb{E}\left[Y^2 \right]  -\mathbb{E}\left[\natlog I_{n-1}(2XY)\right],
\end{align}
where \eqref{cond_entr1} follows from \eqref{channel-law}. In what follows, we will compute the 5 expectations in \eqref{cond_entr1}.

The third and fourth terms in \eqref{cond_entr1} can be readily obtained using \eqref{Rayleigh}
\begin{equation}\label{ElogX}
\mathbb{E}\left[\natlog X\right]= \frac{1}{2} \left(\natlog \rho +\psi(1)\right),
\end{equation}
\begin{equation}\label{EX2}
\mathbb{E} \left[ X^{2}\right] = \rho.
\end{equation}

To compute the second and fifth terms in \eqref{cond_entr1}, we first calculate the output distribution as
\begin{align}\label{output.0}
p_Y(y) 	& = \intop_0^\infty p_{X,Y}(x,y) \, dx \\
\label{output}
  		& = \frac{2y}{\rho \alphaproof^{n-2}} e^{-\frac{y^2}{\rho+1}}\left(1-e^{-\alphaproof y^2}\sum_{k=0}^{n-2} \frac{(\alphaproof y^2)^k}{k!}\right),
\end{align}
where the joint distribution $p_{X,Y}(x,y)$ can be expressed using \eqref{channel-law} and \eqref{Rayleigh} as
\begin{align}\label{joint} 
p_{X,Y}(x,y) 
&= \frac{4}{\rho} \frac{y^{n}}{x^{n-2}}\exp\left(-\frac{x^{2}+\alphaproof y^{2}}{\alphaproof}\right) I_{n-1} (2xy),
\end{align}
with
\begin{align}\label{alphaproof}
\alphaproof\triangleq \frac{\rho}{\rho+1}<1,
\end{align}
and where \eqref{output} can be obtained using a symbolic integration software. Using \eqref{output}, we obtain (using a symbolic integration software)
\begin{equation}\label{ElogY}
\mathbb{E} \left[\natlog Y\right]= \frac{1}{2}\left( \alphaproof \Phi(\alphaproof,1,n) + \psi(n)\right),
\end{equation}
where $\psi(n)$ is the digamma function, $\Phi(\alphaproof,1,n)$ is given by \eqref{PHI}. The second moment of the output distribution is obtained directly from the channel input-output relation \eqref{channel_model}, yielding
\begin{equation}\label{EY2}
\mathbb{E} \left[Y^{2} \right]=\rho + n.
\end{equation}
Substituting \eqref{ElogX}, \eqref{EX2}, \eqref{ElogY} and \eqref{EY2} into \eqref{cond_entr1}, we have
\begin{align}
\nonumber
h_{Y|X}(\rho)=& -\natlog 2 -\frac{n}{2}\alphaproof \Phi(\alphaproof,1,n) -\frac{n}{2}\psi(n) + \frac{n-1}{2}\\
\label{cond_entr_2}
&\cdot \left(\natlog \rho + \psi(1)\right)  + 2\rho+n - h_{Y|X}^{(6)}(\rho),
\end{align}
where
\begin{align}
h_{Y|X}^{(6)}(\rho) \triangleq \intop_0^\infty \intop_0^\infty p_{X,Y}(x,y) \, \natlog \left[I_{n-1}(2xy)\right] dx \, dy.
\end{align}

The last step is to compute the term $h_{Y|X}^{(6)}(\rho)$, which using \eqref{joint} can be expressed as
\begin{align}
\nonumber
h_{Y|X}^{(6)}(\rho)
= & \frac{4}{\rho} \intop_0^\infty \intop_0^\infty \frac{y^{n}}{x^{n-2}}\exp\left(-\frac{x^{2}+\alphaproof y^{2}}{\alphaproof}\right) \\
& \quad \cdot I_{n-1} \left(2xy\right)  \natlog \left[I_{n-1}(2xy)\right] dx \, dy.
\end{align}
We then make the change of variables $\xi = 2xy$, $\eta=y^2$, with the Jacobian $\partial(x,y)/\partial(\xi,\eta)=(4y^2)^{-1}$, yielding
\begin{align}
\nonumber
h_{Y|X}^{(6)}(\rho) = & \frac{2^{n-2}}{\rho} \intop_0^\infty  I_{n-1} (\xi)  \natlog  \left[I_{n-1}(\xi)\right] \\
\label{4thterm_1}
&\quad \cdot \intop_0^\infty \left(\frac{\eta}{\xi} \right)^{n-2} \exp \left(-\frac{\xi^2}{4\eta\alphaproof}-\eta \right) d \eta \,d \xi.
\end{align}
The integration over $\eta$ can be performed analytically, yielding
\begin{align}
\nonumber
\intop_0^\infty \left( \frac{\eta}{\xi} \right)^{n-2} &\exp \left(-\frac{\xi^2}{4\eta \alphaproof}-\eta \right) \, d \eta  \\
\label{4thterm_2}
	&=2^{2-n} \alphaproof^{(1-n)/2} \, \xi \,K_{n-1}\left( \frac{\xi}{\alphaproof^{1/2}}\right),
\end{align}
where $K_{n}(x)$ is the modified Bessel function of the second kind
of order $n$. Using \eqref{4thterm_2} in \eqref{4thterm_1} gives
\begin{align}
\nonumber
h_{Y|X}^{(6)}(\rho) = & \frac{\alphaproof^{(1-n)/2}}{\rho} \intop_0^\infty  \xi K_{n-1}\left( \frac{\xi}{\alphaproof^{1/2}}\right)\\
\label{4thterm_3}
&\qquad\qquad \cdot I_{n-1} (\xi)  \natlog  \left[I_{n-1}(\xi)\right]  \,d \xi\\
\label{4thterm_4}
					= & \frac{\alphaproof^{(1-n)/2}}{\rho} F_n(\rho).
\end{align}
The proof is completed by using \eqref{4thterm_4} in \eqref{cond_entr_2}, the definition of $\alphaproof$ in \eqref{alphaproof}, and by returning to logarithm base 2.

\section{Proof for Lemma~\ref{Lemma_HY}}\label{sec:asymp}

From \eqref{output}, it follows that the output entropy can then be expressed as\footnote{Similarly to Appendix~\ref{Lemma.App}, the results in this proof are in nats.}
\begin{align}\label{E0}
h_{Y}(\rho) = \log\left(\frac{\rho\alphaproof^{n-2}}{2}\right)-\mathbb{E}\left[\natlog Y\right]+\frac{1}{\rho+1}\, \mathbb{E}\left[ Y^{2}\right] + h_{Y}^{(4)}(\rho),
\end{align}
where $\alphaproof$ is given by \eqref{alphaproof},
\begin{align}
h_{Y}^{(4)}(\rho)
\label{l2.p.2}
&\triangleq \intop_0^\infty p_{X}(x) \intop_0^\infty p_{Y|X}(y|x) \, g_{Y}^{(4)}(y) \,dy\,dx \\
&= \intop_0^\infty p_{Y}(y) \, g_{Y}^{(4)}(y) \,dy,
\end{align}
where $p_Y(y)$ is given by \eqref{output} and
\begin{align}
g_{Y}^{(4)}(y) &\triangleq  -\natlog f(\alpha y^2)  \\
              f(z) &\triangleq 1-e^{-z}\sum_{k=0}^{n-2} \frac{(z)^k}{k!}.
\end{align}
Notice that from its definition it follows that the function $f(z)$ is confined to the interval $0  \leq  f(z) \leq 1$. We shall now prove that $h_{Y}^{(4)}(\rho)$ decays as $O\left[\rho^{-1}\right]$ or faster when $\rho \to \infty$.
Indeed, one has
\begin{align}
\label{h.final.1}
h_{Y}^{(4)}(\rho)  &= - \intop_0^\infty \frac{2y}{\rho\alphaproof^{n-2}} \, e^{-\frac{y^2}{\rho+1}} \,f(\alphaproof y^2) \natlog f(\alphaproof y^2) \,dy \\
                   &= - \frac{1}{\rho\alphaproof^{n-1}} \, \intop_0^\infty  e^{-z/\rho} \,f(z) \natlog f(z) \,dz.
\end{align}
Next, one notices that $h_{Y}^{(4)}(\rho)$ is positive and can be upper-bounded as follows
\begin{align}
\label{h.final.3}
h_{Y}^{(4)}(\rho) &\leq  \frac{1}{\rho\alphaproof^{n-1}} \intop_0^\infty  \left(- \,f(z) \natlog f(z) \right)  \,dz  \\
                  &\triangleq   \frac{C}{\rho\alphaproof^{n-1}}.
\end{align}
It is therefore only left to prove that the integral converges, i.e., that the constant $C$ is finite. This can be done as follows:
\begin{align*}
C &= \intop_0^\infty  \left(- \,f(z) \natlog f(z) \right)   \,dz \\
   &\leq \intop_0^\infty  \left(1- f(z)\right) \,dz \\
&= \intop_0^\infty  e^{-z} \, \sum_{k=0}^{n-2} \frac{z^k}{k!} \, dz  \\
    &=n-1 \\
	&< \infty,
\end{align*}
where in the second line we have used an inequality $-x\ln x \leq (1-x)$, $x\in (0,1]$. Therefore, asymptotically $h_{Y}^{(4)}(\rho)$ decays not slower than $1/\rho$.

The asymptotic expression for the output entropy can be written by combining \eqref{h.final.3}, \eqref{ElogY}, \eqref{EY2} and \eqref{E0}, which yields
\begin{equation}
h_{Y}(\rho)=\frac{1}{2}\natlog \rho + 1 - \frac{\psi(1)}{2}-\natlog 2 + O \left[\rho^{-1}\right].
\label{Hy-asymp}
\end{equation}
The proof is completed by returning to logarithm base 2.

\section{Proof of the asymptotically vanishing rate loss}\label{sec:rateloss}
Here we shall prove that an input distribution bounded (truncated) away from zero gives the same results as Theorem \ref{theorem1} in the limit of large average power $\sigma_S \to \infty$. To this end, consider a system where the transmitted amplitudes $X$ are drawn from a Rayleigh distribution with PDF given in \eqref{Rayleigh}. Let us now introduce a threshold $\cox$ of amplitudes realisations below which our channel law model is expected to be inapplicable. Let us now introduce an alternative system where the symbols $\tilde{X}$ are drawn from a ``truncated'' Rayleigh distribution with PDF
\begin{align}\label{Xo.1}
p_{\tilde{X}}(x) = \frac{1}{1-\eta} \, p_{X}(x) \, H(x - \cox),\qquad x \in [\cox,\infty),
\end{align}
where $H(x - \cox)$ is the Heaviside step function, and $\eta$ is defined as
\begin{align}
\eta &\triangleq \mathbb{P} \left[X < \cox\right]. \label{Xo.2}
\end{align}
This probability can be expressed as
\begin{align}
\eta     &=\intop_{0}^{\cox} p_{X}(x) \, dx \label{Xo.3}\\
     &= \frac{2}{\sigma_{S}^{2}} \intop_{0}^{\cox} x \exp \left(-\frac{x^2}{\sigma_{S}^{2}}\right) \, dx \label{Xo.4}\\
     &= 1 - \exp \left(-\frac{{\cox}^{2}}{\sigma_{S}^{2}}\right). \label{Xo.5}
\end{align}
As discussed in Sec.~III-A and Sec.~IV, the threshold $\cox$ is a constant, and thus, $\lim_{\sigma_S \to \infty}\eta =0$.

To prove that the rate loss tends to zero, we shall prove that
\begin{equation}
\lim_{\sigma_{s} \to \infty} \left[I_{X,Y}-I_{\tilde{X},\tilde{Y}} \right]
= 0
\end{equation}
or equivalently,
\begin{align}
\label{diff_entropy_1}
\lim_{\sigma_{s} \to \infty} \left[h_{Y} - h_{\tilde{Y}}\right] = 0
\end{align}
and
\begin{align}
\label{diff_entropy_2}
\lim_{\sigma_{s} \to \infty} \left[h_{\tilde{Y}|\tilde{X}} - h_{Y|X}\right] = 0.
\end{align}

To prove \eqref{diff_entropy_1}, we have the following:
\begin{align}
p_{\tilde Y}(y) &= \intop_0^\infty p_{Y|X}(y|x) \, p_{\tilde X}(x) dx \\
                &= \frac{1}{1-\eta} \intop_{\cox}^\infty p_{Y|X}(y|x)\,p_X(x) \, dx \\
                &\leq \frac{1}{1-\eta} \intop_{0}^\infty p_{Y|X}(y|x)\,p_X(x) \, dx \\
                &= \frac{1}{1-\eta} \, p_Y(y).
\end{align}
The Kullback-Leibler divergence (relative entropy) between the distributions $p_{\tilde Y} (y)$ and $p_Y (y)$ is defined as
\begin{align}
D \left(p_{\tilde Y} (y)\,\|\, p_Y (y) \right)
&\triangleq \mathbb{E} \left[ \log \frac{p_{\tilde Y} (Y)}{p_Y (Y)} \right] \\
&= \intop_{0}^{\infty} p_{\tilde Y} (y) \log \frac{p_{\tilde Y} (y)}{p_Y (y)} \, dy \\
&\leq\log \frac{1}{1-\eta} \intop_{0}^{\infty} p_{\tilde Y} (y) \, dy \\
&=-\log \, (1-\eta) \\
&=\frac{\cox^2}{\sigma_{S}^{2}}.\label{rel.ent.1}
\end{align}
Using the nonnegativity property of the relative entropy together with \eqref{rel.ent.1}, we obtain
\begin{equation} \label{KLD_2}
\lim_{\sigma_{S} \to \infty} D \left(p_{\tilde Y} (y)\,\|\, p_Y (y) \right) = 0.
\end{equation}
Using the fact that the relative entropy is zero if and only if $p_{\tilde{Y}} (y) =  p_Y (y)$ \emph{almost everywhere} \cite[Theorem~8.6.1]{CoverThomas}, we conclude that \eqref{diff_entropy_1} is fulfilled since the integrands in the differential entropy integrals differ on a set with measure zero.

Let us now turn to the first conditional differential entropy in \eqref{diff_entropy_2}, for which we have
\begin{align}
h_{\tilde{Y}|\tilde{X}}
&\triangleq - \intop_{0}^{\infty}\intop_{0}^{\infty} p_{Y|X}(y|x) \, p_{\tilde{X}}(x) \log p_{Y|X}(y|x) \, dx dy \\
&= \intop_{0}^{\infty}  p_{\tilde X}(x) \, g(x) \,dx,
\end{align}
where  
\begin{equation}
\label{g_func}
g(x) \triangleq - \intop_{0}^{\infty} p_{Y|X}(y|x) \log p_{Y|X}(y|x) \,dy
\end{equation}
represents the conditional differential entropy of $p_{Y|X}(y|x)$, and $p_{Y|X}(y|x)$ is given by the noncentral chi-distribution (\ref{channel-law}).

Using \eqref{Xo.1}, the conditional differential entropy $h_{\tilde{Y}|\tilde{X}}$ can be expressed as
\begin{align}
h_{\tilde{Y}|\tilde{X}} &= \frac{1}{1-\eta} \intop_{\cox}^{\infty} p_{X}(x) \, g(x) \, dx,\\
&=\frac{h_{Y|X}}{1-\eta} - \frac{1}{1-\eta}  \intop_{0}^{\cox} g(x) \, p_{X}(x)\,dx. \label{cond.ent.1}
\end{align}

The first term in the r.h.s. of (\ref{cond.ent.1}) tends to the conditional entropy of the untruncated distribution.  We shall now prove that the last (integral) term in \eqref{cond.ent.1} tends to zero when $\sigma_{S} \to \infty$. We note that according to \eqref{Rayleigh} the input distribution $p_{X}(x)$ tends to zero uniformly in the interval $[0,\cox]$ as $\sigma_{S} \to \infty$. Then, according to the bounded convergence theorem, in order to prove that integral term in \eqref{cond.ent.1} is asymptotically vanishing, it is sufficient to prove that the function $g(x)$ remains bounded within the interval $[0,\cox]$. We shall do so by providing separate upper and lower bounds for this function.

The upper bound for $g(x)$ can be obtained by considering a relative entropy between the channel law $p_{Y|X}(y|x)$ and an auxiliary distribution $p_{Y}^{\diamond}(y)$ supported on $[0,\infty)$. The nonnegativeness of the relative entropy immediately provides an upper bound for the differential entropy \eqref{g_func}, namely, 
\begin{align}
g(x) \leq - \, \mathbb{E} \left[\, \log p_{Y}^{\diamond}(Y) \,\right] = -\intop_{0}^{\infty}p_{Y|X}(y|x) \log p_{Y}^{\diamond}(y) \,dy.
\end{align}
Choosing a half-Gaussian distribution $p_{Y}^{\diamond}(y) = \left(2/\sqrt{\pi}\right) \exp \left(-y^2\right)$ immediately gives an upper bound  $g(x) \leq \mathbb{E}\left[Y^{2}\right] - \log \left(2/\sqrt{\pi}\right)$. The second moment for the noncentral chi distribution is readily available, e.g., from \eqref{channel_model}, leading to the following upper bound:
\begin{equation}
\label{g.upper}
g(x) \leq x^2 +n\,\sigma_N^2 + \log \frac{\sqrt{\pi}}{2}.
\end{equation}
Note that this upper bound is bounded inside an arbitrary finite interval $[0,\cox]$.

Establishing a lower bound for $g(x)$ is slightly more involved. The first step is to transform the noncentral chi distribution into a noncentral chi-squared distribution by making the following change of variable in the integral (\ref{g_func}): $z=2y^2/\sigma^2_N$. Introducing the additional notation $\lambda=2x^2/\sigma_N^2$ and $n=k/2$, where $k$ is a number of degrees of freedom of noncentral chi-squared distribution, we obtain \begin{equation}
\label{nc_chi2}
p_{Z|\Lambda}(z|\lambda) = \frac{1}{2} \left(\frac{z}{\lambda}\right)^{(k-2)/4}\exp\left(-\frac{z+\lambda}{2}\right) I_{(k-2)/2}(\sqrt{\lambda z})
\end{equation}
with $z \in [0, \infty)$. We can now express $g(x)$ in \eqref{g_func} as an average with respect to the noncentral chi-squared distribution:
\begin{align}
g(\lambda) &=- \intop_{0}^{\infty} p_{Z|\Lambda}(z|\lambda) \log \left[\frac{2^{3/2}\, z^{1/2}}{\sigma_N}\, p_{Z|\Lambda}(z|\lambda)\right] \,dz \\
		 &= g^{(1)}(\lambda) + g^{(2)}(\lambda) + \frac{3}{2} \log 2 - \log \sigma_N,\label{g.def}
\end{align}
where we have introduced two functions: $g^{(1)}(\lambda)$, which represents the differential entropy of the noncentral chi-squared distribution $p_{Z|\Lambda}(z|\lambda)$, i.e.,
\begin{equation} \label{gg_1}
 g^{(1)}(\lambda) \triangleq -\intop_{0}^{\infty} p_{Z|\Lambda}(z|\lambda) \log p_{Z|\Lambda}(z|\lambda)\,dz,
\end{equation}
and $g^{(2)}(\lambda)$, which stands for minus half of the so-called \emph{expected-log}, i.e.,
\begin{equation} \label{gg_2}
g^{(2)}(\lambda) \triangleq  - \frac{1}{2} \, \mathbb{E} \left[ \,\log Z \,\right].
\end{equation}
The motivation for the above transformation stems from the fact that it has been proven in \cite{yyu11} that the noncentral chi-squared distribution function \eqref{nc_chi2} is \emph{log-concave} (i.e., log of $p_{Z|\Lambda}(z|\lambda)$ is concave) if the number of degrees of freedom $k\geq 2$, i.e., $n\geq 1$, which is always the case. On the other hand, the differential entropy of any log-concave distribution function can be lower-bounded as \cite[Theorem 3]{mk17}
\begin{align}
g^{(1)}(\lambda) &\geq \log \left(2 \sqrt{\var\left[Z\right]}\right) =\frac{1}{2} \log \left(k+2\lambda\right) + \frac{3}{2} \log 2.\label{g1.lb}
\end{align}

Finally, let us now provide a lower bound for $g^{(2)}(\lambda)$ in \eqref{gg_2}. This can be obtained by applying Jensen's inequality:
\begin{align}
g^{(2)}(\lambda)  & \geq  -\frac{1}{2} \log \mathbb{E} \left[Z\right]  =  -\frac{1}{2} \log \left(k+\lambda\right).\label{g2.lb}
\end{align}
Combining \eqref{g.def}, \eqref{g1.lb}, and \eqref{g2.lb}, and returning to the original notation, we obtain
\begin{equation}
\label{g.lower}
g(x) \geq \frac{1}{2} \log \left(\frac{2 x^2 +n\sigma_N^2}{x^2+n\sigma_N^2} \right) - \log\,\sigma_{N} + 3 \log 2.
\end{equation}
This lower bound on $g(x)$ is bounded inside an arbitrary finite interval $x \in [0,\cox]$. Thus, the function $g(x)$ in the integral \eqref{cond.ent.1} is uniformly bounded via \eqref{g.upper} and \eqref{g.lower} in $[0,\cox]$. Since in the asymptotic limit $\sigma_S \to \infty$ one has $\eta \to 0$ and $p_{X}(x) \to 0$ from \eqref{cond.ent.1}, it follows that \eqref{diff_entropy_2} is fulfilled as well, which concludes the proof.

\section*{Acknowledgments}\label{Sec:Ack}

The authors would like to thank the anonymous reviewers for their valuable comments.

\begin{IEEEbiographynophoto}
{Nikita~A.~Shevchenko} obtained the BSc and MSc degrees, both in applied physics, from Donetsk National University, Ukraine, in 2008 and 2009, respectively, where he graduated \emph{summa cum laude}. He then enrolled on a PhD programme in condensed matter physics at Donetsk O.~O.~Galkin Institute for Physics and Engineering of National Academy of Sciences of Ukraine, focusing on theoretical investigations in nonlinear optics. During his studies he spent some time as an academic visitor at the University of Exeter, where he was involved in a research project with the Electromagnetic and Acoustic Materials Research group. In October 2013, he joined the Optical Networks Group at University College London (UCL), London, United Kingdom, where he is currently working toward the PhD degree. His main research interests lie in nonlinear optics, photonics, optical communications and information theory.

\end{IEEEbiographynophoto}

\begin{IEEEbiographynophoto}
{Stanislav Derevyanko} received his Ph.D. degree in theoretical physics
in 2001 from Institute for Radiophysics and Electronics, Kharkiv, Ukraine.
From 2002 to 2007, he was twice a Postdoctoral Researcher fellow at  Photonics Research Group at 
Aston University, Birmingham UK and from 2007 to 2012 he was an EPSRC Advanced Fellow at Nonlinearity
And Complexity Research Group also at Aston University. From 2013 to 2015 he was a Marie Curie visiting Fellow 
at Weizmann Institute of Science, Israel and in 2015 he joined the Department  of Electrical and Computer 
Engineering at Ben Gurion University of the Negev, Beer Sheva, Israel, 

His research interests include nonlinear optics, optical telecommunications and information theory.
\end{IEEEbiographynophoto}

\begin{IEEEbiographynophoto}
{Jaroslaw E. Prilepsky} received his Master's degree (M.E.) in Theoretical Physic (Degree with first class honours) from V. Karazin Kharkiv National University, Ukraine, in 1999, and Ph.D. degree in Theoretical Physics from the B. Verkin Institute of Low Temperature Physics and Engineering, Kharkiv, Ukraine, in 2003, studying nonlinear excitations in low-dimensional systems. From 2003 till 2010 he was a Research Fellow in B. Verkin Institute of Low Temperature Physics and Engineering, and in 2004 he was a Visiting Fellow in Nonlinear Physics Center, Research School of Physical Sciences and Engineering, Australian National University, Canberra, Australia. In 2010-2012 he became a Research Associate in Nonlinearity and Complexity Research Group, Aston University, UK, and from 2012 till now he has been a Research Associate in Aston Institute of Photonics Technologies, Aston University. His current research interests include (but not limited to) optical transmission systems and networks, nonlinearity mitigation methods, nonlinear Fourier-based optical transmission methods, soliton usage for telecommunications, information theory, and methods for optical signal processing. His is the author of more than 60 journal papers and conference contributions in the fields of nonlinear physics, solitons, nonlinear signal-noise interaction, optical transmission, and signal processing.
\end{IEEEbiographynophoto}

\begin{IEEEbiographynophoto}
{Alex Alvarado} (S'06--M'11--SM'15) was born in Quell\'{o}n, on the island of Chilo\'{e}, Chile. He received his Electronics Engineer degree (Ingeniero Civil Electr\'{o}nico) and his M.Sc. degree (Mag\'{i}ster en Ciencias de la Ingenier\'{i}a Electr\'{o}nica) from Universidad T\'{e}cnica Federico Santa Mar\'{i}a, Valpara\'{i}so, Chile, in 2003 and 2005, respectively. He obtained the degree of Licentiate of Engineering (Teknologie Licentiatexamen) in 2008 and his PhD degree in 2011, both of them from Chalmers University of Technology, Gothenburg, Sweden.

Dr. Alvarado is an assistant professor at the Signal Processing Systems (SPS) Group, Department of Electrical Engineering, Eindhoven University of Technology (TU/e), The Netherlands. During 2014--2016, he was a Senior Research Associate at the Optical Networks Group, University College London, United Kingdom. In 2012--2014 he was a Marie Curie Intra-European Fellow at the University of Cambridge, United Kingdom, and during 2011--2012 he was a Newton International Fellow at the same institution. Dr. Alvarado's current research is funded in part by the Netherlands Organisation for Scientific Research (NWO) via a VIDI grant, as well as by the European Research Council (ERC) via an ERC Starting grant.

Dr. Alvarado is a recipient of the 2009 IEEE Information Theory Workshop Best Poster Award, the 2013 IEEE Communication Theory Workshop Best Poster Award, and the 2015 IEEE Transaction on Communications Exemplary Reviewer Award. He received the 2015 Journal of Lightwave Technology best paper award, honoring the most influential, highest-cited original paper published in the journal in 2015. Dr. Alvarado is a senior member of the IEEE and an associate editor for IEEE Transactions on Communications (Optical Coded Modulation and Information Theory). Since 2018, he also serves in the OFC subcommittee Digital and Electronic Subsystems (S4). His general research interests are in the  areas of digital communications, coding, and information theory.
\end{IEEEbiographynophoto}

\begin{IEEEbiographynophoto}
{Polina Bayvel} received her BSc (Eng) and PhD degrees in Electronic \& Electrical Engineering from University of London, UK, in 1986 and 1990, respectively. In 1990, she was with the Fiber Optics Laboratory, General Physics Institute, Moscow (Russian Academy of Sciences), under the Royal Society Postdoctoral Exchange Fellowship. She was a Principal Systems Engineer with STC Submarine Systems, Ltd., London, UK, and Nortel Networks (Harlow, UK, and Ottawa, ON, Canada), where she was involved in the design and planning of optical fibre transmission networks. During 1994-2004, she held a Royal Society University Research Fellowship at University College London (UCL), and in 2002, she became a Chair in Optical Communications and Networks. She is the Head of the Optical Networks Group (ONG), UCL which she also set up in 1994. She has authored or co-authored more than 300 refereed journal and conference papers. Her research interests include wavelength-routed optical networks, high-speed optical transmission, and the study and mitigation of fibre nonlinearities.

Prof. Bayvel is a Fellow of the Royal Academy of Engineering (FREng), the Optical Society of America, the UK Institute of Physics, and the Institute of Engineering and Technology. She was the recipient of the Royal Society Wolfson Research Merit Award (2007-2012), 2013 IEEE Photonics Society Engineering Achievement Award, and the 2014 Royal Society Clifford Patterson Prize Lecture and Medal. In 2015 she and 5 members of ONG received the Royal Academy of Engineering Colin Campbell Mitchell Award for their pioneering contributions to optical communications technology.
\end{IEEEbiographynophoto}

\begin{IEEEbiographynophoto}
{Sergei Turitsyn} received the Graduate degree from the Department of Physics,
Novosibirsk University, Novosibirsk, Russia, in 1982 and the Ph.D. degree in
theoretical and mathematical physics from the Institute of Nuclear Physics,
Novosibirsk, in 1986. In 1992 he moved to Germany, first as a Humboldt
Fellow, and then working in the collaborative projects with Deutsche Telekom.
Currently, he is the Director of the Aston Institute of Photonic Technologies. He
received the Royal Society Wolfson Research Merit Award in 2005. In 2011,
he received the European Research Council Advanced Grant,  in 2014 he
received Lebedev medal by the Rozhdestvensky Optical Society, and in 2016
Aston 50th Anniversary Chair medal. He is a fellow of the Optical Society of
America and the Institute of Physics.
\end{IEEEbiographynophoto}

\end{document}